\documentclass{JHEP3}

\usepackage{epsfig}
\usepackage{amssymb}
\usepackage{amscd}
\usepackage{amstext}
\usepackage{graphics}

\newcommand{\bea}{\begin{eqnarray}}
\newcommand{\eea}{\end{eqnarray}}
\newcommand{\bean}{\begin{eqnarray*}}
\newcommand{\eean}{\end{eqnarray*}}

\newcommand{\hs}{\hspace{0.3 cm}}
\newcommand{\nf}{\text{\large{n}}^{\hspace{-0.03 cm} \text{\tiny{F}}}}
\newcommand{\ig}{\includegraphics}
\newcommand{\rb}{\raisebox}
\newcommand{\tsg}[2]{\rb{- #1}{\rule{0pt}{#2}}}

\newcommand{\ag}{\alpha}
\newcommand{\bg}{\beta}
\newcommand{\cg}{\gamma}
\newcommand{\dg}{\delta}

\def\IZ{\mathbb{Z}}
\def\IF{\mathbb{F}}

\def\IC{\mathbb{C}}
\def\IP{\mathbb{P}}

\def\mQ{\mathcal{Q}}

\def\be{\begin{equation*}}
\def\ee{\end{equation*}}
\def\beq{\begin{equation}}
\def\eeq{\end{equation}}

\newcommand{\fref}[1]{Figure~\ref{#1}}

\preprint{MIT-CTP-3505}

\title{New results on superconformal quivers}

\author{Sergio Benvenuti$^1$, $\,$ Amihay Hanany$^2$\\

\vspace{0.3 cm}

1. \parbox[t]{6in}{Scuola Normale Superiore, Pisa,\\ 
                   and INFN, Sezione di Pisa, Italy.}

\vspace{0.5 cm}

2. \parbox[t]{6in}{Center for Theoretical Physics,\\
                   Massachusetts Institute of Technology,\\ 
                   Cambridge, MA 02139, USA.}\\

\email{sergio.benvenuti@sns.it, hanany@mit.edu}
}

\abstract{All superconformal quivers are shown to satisfy the relation c = a and are thus good candidates for being the field theory living on D3 branes probing CY singularities. We systematically study 3 block and 4 block chiral quivers which admit a superconformal fixed point of the RG equation. Most of these theories are known to arise as living on D3 branes at a singular CY manifold, namely complex cones over del Pezzo surfaces. In the process we find a procedure of getting a new superconformal quiver from a known one. This procedure is termed ``shrinking" and, in the 3 block case, leads to the discovery of two new models. Thus, the number of superconformal 3 block quivers is 16 rather than the previously known 14. We prove that this list exausts all the possibilities. We suggest that all rank 2 chiral quivers are either del Pezzo quivers or can be obtained by shrinking a del Pezzo quiver and verify this statement for all 4 block quivers, where a lot of ``shrunk'' del Pezzo models exist.}
\begin{document}

\newpage

\section{Introduction}
\label{section_introduction}
Placing a stack of $N$ coincident D3 branes at the singular point of a Calabi-Yau cone, it is possible to geometrically engineer a large class of superconformal gauge theories living in four dimensions. Since the massless excitations in the decoupling limit have their origin in open strings stretched between the D3 branes, all the fields of the gauge theories are in representations of the gauge group with exactly two indices; this implies that the matter content of these gauge theories can be encoded in quiver diagram.

A supersymmetric quiver is a quantum field theory with product gauge group $\Pi_i U(N_i)$ and matter composed by chiral superfields transforming in the bifundamental or adjoint representations. The name comes from the fact that it is possible to represent the field content of the theory with a quiver diagram: each gauge factor $U(N_i)$ corresponds to a node in the diagram, the bifundamental fields correspond to arrows connecting two different nodes and the adjoint fields are arrows starting and ending in the same node. The bifundamental fields transform in the $(\overline{N}_i, N_j)$ representations of the groups $U(N_i)$ and $U(N_j)$, located at the tip and at the tail of the arrow. In this sense $\mathcal{N} = 4$ SYM is one of the simplest supersymmetric quivers, with just one node and three arrows.

Since all the fields are in $2$-index representations of the (in general non simple) gauge group, quiver gauge theories admit a natural one-parameter large $N$ expansion. $N$ is simply the greatest common divisor of the ranks $\{N_i\}$: $N_i = N \, x_i$.

For all the models arising from D3 branes at singularities it is possible to take the near horizon limit \cite{Maldacena:1997re}\cite{morrisonplesser}: there is a gravity dual provided by Type IIB superstring compactified on $AdS_5 \times X_5$, where $X_5$ is the base of the Calabi-Yau cone. Much of the predictive power of the duality resides in the fact that there is a limit in which a description by \emph{weakly coupled} gravity on $AdS_5$ is adequate. This limit corresponds to taking the large $N$ limit and to selecting a particular point in the manifold of 4D conformal field theories. This point is always located in a limit of infinite couplings, for instance in the basic case of $\mathcal{N} = 4$ SYM it correspond to $g_{YM}^2 N = \infty$.

In \cite{henningson} Henningson and Skenderis showed, for this last type of models, the validity of a peculiar relation between ``$c$'' and ``$a$'', the two gravitational central charges of the 4D conformal field theory: they are always equal in the large $N$ limit. This is a consequence of the fact that the dual gravity on $AdS_5$ is weakly coupled. Since the values of the central charges ``$c$'' and ``$a$'' are invariant under marginal deformations, the relation $c = a$ is true for the whole set of 4D CFTs containing the one with a weakly coupled gravity dual.

In this paper we show that the relation $c = a$ holds for any superconformal quiver, without assuming the existence of a brane description of the gauge theory. This result suggests that all superconformal quivers can be engineered from D3 branes probing a singular Calabi-Yau threefold.

Motivated by this observation, we start a classification of superconformal quivers, giving some more evidence to this expectation. We concentrate on the class of models that we find more interesting: completely chiral quivers, i.e. models whose quiver diagram has no bidirectional arrows. We complete the classification of the models that can be Seiberg dualized to a $3$ or $4$ block quiver (the precise definitions are given at the beginning of Section \ref{threeblock}). The well studied theories arising from branes on vanishing del Pezzo surfaces belong to this class.

The stategy of the classification is as follows: we first show that a superconformal fixed point exists only if a highly restrictive condition for the discrete parameters of the quiver is satisfied. This condition is a diophantine equation for the number of arrows between the nodes, and depends parametrically on the number of nodes in each block (it is crucial that the ranks of the gauge groups do not enter in this equation). The superconformal models are in correspondence with the solutions of these equations. Different solutions to the same equation correspond to different Seiberg dual phases of the same gauge theory.

During the process of the classification we discover an interesting operation, that enables one to construct new quivers from known ones. This operation is called ``shrinking'', since at the level of quiver diagrams it consists in replacing a block composed by $n^2$ nodes with a single node. The applicability of the ``shrinking'' procedure is not restricted to chiral quivers, it can be applied to any superconformal quiver. A general geometric interpretation of this procedure should be possible; we conjecture that it is related to Discrete Torsion, i.e. turning on discrete fluxes for RR and NS $3$-form field strengths.

The surprising result of the classification is that \emph{all} $3$ and $4$ block superconformal chiral quivers can be obtained by a del Pezzo quiver (for which the geometric interpretation is well known), shrinking one or two blocks composed by four nodes. This in turn suggests that all quivers whose quiver matrix has rank $2$ can be constructed in this way, without any assumptions about a block structure. If this is true the number of nodes of any rank-$2$ chiral quiver is at most $11$, i.e. the number of nodes of the quiver for the last del Pezzo surface, del Pezzo $8$.

\vspace{0.6 cm}

The paper is organized as follows.

In section \ref{general} we give the general proof of the equality between the central charges, then we discuss Seiberg dualities and the problem of finding the precise form of the superpotential. A comment on the relation $c = a$ along Renormalization Group flows is also made.

The structure of $3$-block quivers without adjoints and bidirectional arrows (chiral) is discussed at lenght in section \ref{threeblock}. Here we show, using a result proved in the appendix, that the diophantine equation classifying all possible models imposes that there are exactly $16$ $3$-block quivers.

In section \ref{quiveroperation} two general operations on superconformal quivers are discussed, the orbifold and the ``shrinking'' operations. In our case of interest (chiral quivers), the ``shrinking'' procedure enables to contruct a lot of new rank $2$ chiral superconformal models, starting from the well known del Pezzo quivers.

In the case of $4$ blocks we show, in section \ref{fourblock}, that the superconformal constraints impose, as for $3$ blocks, that only del Pezzo and shrunk del Pezzo quivers are possible.

\section{General properties of superconformal quivers}\label{general}
In this section we show that all interacting superconformal quivers satisfy the relation $c = a$, by proving a formula that relates the difference between $c$ and $a$ to a weighted sum of the beta functions for the gauge couplings. We then comment on the implications of this formula for Renormalization Group flows driven by superpotential couplings, comparing with supergravity results on the axion-dilaton expectation values. We then discuss general features of quiver gauge theories such as the existence of Duality Trees and the problem of finding the exactly marginal superpotential.

A quivers has the superconformal symmetry if all the couplings sit at a Renormalization Group fixed point, i.e. the beta function for each coupling is zero. Moreover we require this fixed point to be \emph{completely interacting}, in the sense that \emph{all} the couplings are at a non-trivial zero of the beta functions (indeed if a coupling is zero its corresponding beta function vanishes trivially). This property can be seen as a direct consequence of the definition of quivers: if one of the gauge couplings vanishes the corresponding gauge group decouples from the rest of the theory, and the bifundamental fields which are linked to this node ``lose'' one index and become just fundamental. Thus if one of the gauge coupling is free, the model will not be in the class of quiver theories we consider.

We will see that these interacting fixed points are never isolated. There is always a manifold of RG fixed points. Sometimes this manifold contains the free theory. An example of this is $\mathcal{N} = 4$ SYM or, more generally, orbifolds of $\IC^3$. This is the sense in which the requirement of interacting gauge couplings has to be interpreted in these cases: we have to rule out the possibility of a fixed point manifold where one (or more) of the couplings is forced to vanish.

We stress that the relation $c = a$ is quite a remarkable property for 4D conformal field theories; for example the $OPE$ of the stress-energy tensor with itself in general contains other operators (like the Konishi current), but for theories with $c = a$ this OPE is closed, meaning that the singular part of the OPE contains only the identity operator and the stress-energy tensor \cite{Anselmi:1998ms}, \cite{Anselmi:1999bc}, \cite{Anselmi:2000}. We would also like to remark that our arguments use strong coupling properties of these 4D theories and crucially depend on supersymmetry, while the results of \cite{Maldacena:1997re}\cite{morrisonplesser}\cite{henningson}\cite{Anselmi:2000} are valid for non supersymmetric conformal theories as well.

\subsection{Superconformal quiver implies $c = a$}\label{c=a}
In this paragraph we give a detailed analysis of the relation $c = a$ for superconformal quivers. The only ingredients we will use are the fact that the matter is in $2$-index representations and the fact that all gauge couplings are at an interacting RG fixed point.

The two gravitational central charges of a superconformal field theory, $c$ and $a$, are defined as the coefficients of the two possible terms of the trace anomaly. In a background gravitational field $g^{\mu \nu}$ the trace of the energy momentum tensor is:
\beq
\label{defac} \Theta [g^{\mu \nu}] = \frac{c}{16 \pi^2}
(\mathcal{W_{\ag \bg \cg \dg}})^2 - \frac{a}{16 \pi^2}
(\tilde{R}_{\ag \bg \cg \dg})^2 + \partial^{\ag} J_{\ag}\;\;,
\eeq
where $W_{\ag \bg \cg \dg}$ is the Weyl tensor, $\tilde{R}_{\ag \bg \cg \dg}$ is the dual of the Riemann tensor \footnote{An alternative definition is to take the coefficients of the 2 and 3 point function of the energy momentum tensor in flat space. This can be seen by differentiating (\ref{defac}) with respect to the metric and then taking the flat space limit.} and the last term represents scheme dependent total derivatives unimportant for us.

In \cite{Anselmi:1998am}, \cite{Anselmi:1998ys} exact formulae relating the gravitational central charges to the 't Hooft anomalies of the $R$-current have been found: 
\beq\label{AFGJ}
\left\{
\begin{array}{l}
c  =  \frac{3}{32} \left( 3 \, tr R^3 -\, tr R \right)\\
a = \frac{3}{32} \left( 3 \, tr R^3 - 5 \, tr R \right)\;.
\end{array}
\right.
\eeq
Inverting this system we see that $tr R$ is proportional to $c - a$, so the condition $c = a$ is equivalent to $tr R = 0$.

We recall that the traces of the $R$-current, $tr R$ and $tr R^3$, are by definition fermionic traces coming from the triangle anomalies of the $R$-symmetry current with itself or with the energy-momentum tensor $<\!R R R\!>$ and $<\!R T T\!>$. For $tr R^\ag$ there are thus contributions from the gauginos (carrying $r$-charge $1$) and from the fermionic components of the chiral superfields in the representation $\mathcal{R}_M$ (with $r$-charge $r_M$ and dimension $dim [\mathcal{R}_M]$):
\beq\label{traceR}
tr R^{\ag} = \sum_{G\in gauge\,groups}\hspace{-0.4 cm} dim[G]  \, (1)^{\ag} +
\sum_{M\in matter fields} \hspace{-0.4 cm} dim [\mathcal{R}_M] (r_M - 1)^{\ag} \;.
\eeq
The contribution from the gauge fields is always of order $N^2$, since the dimension of the adjoint of $SU(N)$ is $N^2-1$. If there are only fields in 2-index representation also the contribution from the matter fields is $O(N^2)$. Thus, for quivers, $c$, $a$, $tr R$, $tr R^3$ are all of order $N^2$. We are going to show that $tr R$ is $O(1)$ for a supersymmetric quiver gauge theory at an interacting conformal point.

We now need the formula for the beta function of the gauge couplings. In the case of $SU(N)$ the NSVZ beta function (exact to all orders in perturbation theory) is:
\beq\label{NSVZa}
\beta_{\frac{1}{g^2}} = \frac{1}{8 \pi^2} \frac{3 N - \sum_{M} \mu[\mathcal{R}_M]  (1 - \cg_{M}(g))}{1 - g^2 N / 8 \pi^2}\;,
\eeq
where $\mu[\mathcal{R}]$ is the Dinkin index of the representation $\mathcal{R}$, defined by $Tr_{\mathcal{R}}T^a T^b = \mu[\mathcal{R}] \delta^{a\,b}$ (our conventions are such that for the fundamental representation $\mu = \frac{1}{2}$, so for the adjoint of $SU(N)$ $\mu = N$). In order to study the superconformal fixed point it is enough to take in consideration the zeros of the numerator. So we define $\bg$ to be
\beq\label{NSVZb}
\beta =  N + \sum_{M} \mu[\mathcal{R}_M]  (r_M - 1)\;,
\eeq
where we changed variables, from the anomalous dimensions $\cg$ to the $r$-charges, using the relation between the total scaling dimension $D$ of a chiral operator and its $r$-charge
$$D = 1 + \cg /2 = 3 / 2 \; r\,.$$
In the rest of the paper we will always take in consideration just the numerator of the beta functions, and the variables will always be the $r$-charges of the elementary chiral superfields. For quivers the matter resides in the adjoints and in the bifundamentals, so (\ref{NSVZb}) becomes:
\beq
\beta_i =
N_i + \sum_{A \in adj[i]} N_i (r_{A, i} - 1) + \frac{1}{2} \sum_{B
\in bifund[i, j]} N_j (r_{B, i j}-1)\;.
\eeq
Here $N_i$ is the rank of the $i^{th}$ group, $SU(N_i)$. The first sum is on the adjoint multiplets of the node $i$, carrying $r$-charge $r_{A, i}$. The second sum is on all the nodes $j$ different from $i$ and all bifundamentals connecting node $i$ to node $j$, having $r$-charge $r_{B, i j}$.

Let's consider the sum of the previous beta functions, weighted with the ranks of the gauge group $N_i$:
\beq \sum_i N_i \beta_i =
\sum_i N_i^2 + \sum_{i}\sum_{A \in adj[i]} N_i^2 (r_{A, i} - 1) +
\frac{1}{2}\sum_{i, j} \sum_{B \in bifund[i,j]} N_i N_j (r_{B, i
j} - 1) \;.
\eeq
The second sum is in fact a sum over all the adjoint chiral fields of the theory. The third sum counts each bifundamental field exactly two times, so it can be replaced by a sum over all bifundamental fields, without the factor $\frac{1}{2}$:
\beq
\sum_i N_i \beta_i = \sum_i N_i^2 + \sum_{A \in adjoints} dim[A]\,(r_A - 1) + \sum_{B \in bifund} dim[B]\,(r_B - 1) \;.
\eeq
The last expression matches the formula (\ref{traceR}) for $tr R$ in the case of adjoint and bifundamental matter, so we conclude:
\beq\label{fund}
tr R = \sum_i N_i \beta_i \;.
\eeq
Since at an interacting RG fixed point $\beta_i = 0$ for each gauge group, $tr R$ will vanish as well.

The consequence is that for any superconformal quiver $c = a$ in the large $N$ limit.

We stress that this simple proof of the relation $c = a$ is purely field-theoretical and does not rely on any assumption about holography, in particular it does not require the existence of a holographic dual for the superconformal quiver.

The argument above is true in the large $N$ limit, where both $c$ and $a$ are of order $N^2$. There are corrections of order $1$ coming from the fact that the dimension of the representation for the adjoint chiral superfields is $N_i^2 - 1$ instead of $N_i^2$. Another order $1$ correction comes from considering $U(N)$ gauge groups instead of $SU(N)$, in this case the source of the discrepancy would be the number of gauginos.

The same argument would also show that $c = a$ for ``non oriented'' theories: additional $SO(N)/Sp(N)$ gauge groups and matter transforming in the symmetric and in the antisymmetric representation. For ``non oriented'' models there are corrections of order $N$, not just of order $1$.

We would like to point out that we didn't make any assumption about the superpotential of the theory. However this is an important
issue: we will see that it is the superpotential that ensures the possibility of flowing to the manifold of IR fixed points, giving thus strong support to the idea of the existence of the non trivial superconformal theories. Moreover we will see that it is precisely a non zero superpotential that implies the existence of marginal directions for the IR quivers.

\subsection{A comment on RG flows and the string coupling}
In this paragraph\footnote{This subsection is not directly related with the following part of the paper, that can be read independently.} we would like to make some comments on formula (\ref{fund}):
$$   tr R = \sum_i N_i \, \beta_i \,.$$
This relation has been obtained without the requirement of being at criticality, so it can be used to study non conformal supersymmetric quivers as well. Let's consider what happens along a Renormalization Group flow between two superconformal (interacting) quivers driven by a superpotential coupling. Both the UV and IR fixed point are superconformal theories with $tr R = 0$.

It is possible to construct a lot of examples of this kind.

For instance one can take $\mathcal{N} = 4$ SYM and add a mass term for one of the three adjoint superfields. The IR fixed point is a $\mathcal{N} = 1$ theory with two adjoints (of dimension $3/4$) and a quartic superpotential. Another well studied example starts from the $\mathcal{N} = 2$ orbifold $\IC \times \IC^2 / \IZ_2$, called the $\hat{A}_1$ model. If a suited mass term for the adjoints is added \cite{klebanovwitten}, this model flows in the IR to the conifold: a quiver with two gauge groups of the same rank, four bifundamentals and no adjoints. Also in this case the surviving bifundamental matter has dimension $3/4$ (or $r$-charge $1/2$).

Since $tr R = 0$ at the two ends of the RG flow it is natural to ask if $tr R = 0$ along all the flow. However outside criticality $r$-charges and formulae like (\ref{AFGJ}) depend on the renormalization scheme used. So the question should be if there is a ``natural'' renormalization scheme leading to $tr R = 0$ along all the flow.

\vspace{0.2 cm}

A hint toward the identification of this ``natural scheme'' comes from string theory. There is in fact a relation between the string coupling $g_s$ and the gauge couplings of the quiver $g_i$:
\beq\label{gstring}
\frac{N}{4 \pi g_s} = \sum_i  \frac{N_i}{g^2_i}\,.
\eeq
Considering the variation of (\ref{gstring}) with respect to a renormalization scale $\mu$, we see that $tr R$ can be interpreted as the beta function for the string coupling:
\beq \label{betagstring}
\frac{\partial}{\partial \ln \mu}\,\frac{N}{4 \pi g_s} =
\sum_i N_i \frac{\partial}{\partial \ln \mu}\,\frac{1}{g^2_i} = \sum_i N_i \beta_i = tr R\,.
\eeq
In this equation we are forgetting about the denominators of the NSVZ beta functions. The idea is that in the ``natural'' renormalization scheme the denominators are absent. For instance this is the case in the well studied Duality Cascade of Klebanov and Strassler \cite{klebanovstrassler}, were the quantities calculated in field theory and in string theory match precisely if the denominators are not taken in consideration.  

In this sense the interpretation of (\ref{betagstring}) is that $\sum_i N_i \beta_i = tr R$ is the natural definition of $tr R$ away from criticality. The consequence is that, if $g_s$ remains constant, $tr R$ vanishes along all the flow. This is clearly expected: we are considering flows driven by superpotential terms, so, even if the $r$-charges are varying along the flow, the beta functions for the $gauge$ couplings should remain zero and, by (\ref{betagstring}), $tr R$ should remain zero as well. The beta functions for the superpotential couplings that drive the flow are clearly different from zero, but these do not enter in our formula (\ref{fund}) for $tr R$. 

If instead we consider holographic flows with varying $g_s$, equation (\ref{betagstring}) implies that the difference between the values of $g_s$ at the UV and at the IR is given by the integral of $tr R$:
\beq
\int_0^\infty tr R \; d\ln\mu = \frac{N}{4 \pi}\Delta g_s^{-1}\;. 
\eeq
Let's consider the situation in which the entire RG flow can be described by Type IIB supergravity. This should correspond to having non trivial superconformal quivers both in the UV and IR fixed point. In these case the dilaton can be set to a constant, leading to $tr R = 0$ along the entire flow. 

\vspace{0.2 cm}

On the gauge theory side the recent proposal of Kutasov \cite{kutasov03} is related to this discussion. In \cite{kutasov03}\cite{kutasovschwimmer}\cite{intriligator04} it is found, by a different path, that the ``natural'' scheme is the one in which the couplings are identified with lagrange multipliers implementing the $a$-maximization of \cite{intriligator03} outside conformality. The anomalous dimensions (or the $r$-charges) are given by an explicit formula that interpolates between the UV and the IR points \cite{kutasov03} and in this scheme the denominators of the beta functions are absent. We can thus trust the relations of \cite{Anselmi:1998am}, \cite{Anselmi:1998ys}
\beq\label{AFGJb}
\left\{
\begin{array}{l}
c = \frac{3}{32} \left( 3 \, tr R^3 -\, tr R \right)\\
a = \frac{3}{32} \left( 3 \, tr R^3 - 5 \, tr R \right)
\end{array}
\right.
\eeq
along the whole flow, where for the $r$-charges the interpolating formulae of \cite{kutasov03} have to be used. However here a problem occurs: if we take for the definition a $a$ the one used in \cite{kutasov03} (formula (\ref{AFGJb}) plus Lagrange multipliers terms) and for $c$  just formula (\ref{AFGJb}), we see that the relation $c = a$ is spoiled outside conformality, since the Lagrange multipliers terms are proportional to the superpotential beta-function, which is non-zero outside between the UV and IR fixed points. However in \cite{kutasov03} it is noticed that using just (\ref{AFGJb}) the central charge $a$ still satisfies the $a$-theorem. 

There are thus two possibilities: either one takes only formula (\ref{AFGJb}) for the central charges outside criticality, either one adds the lagrange multipliers terms both to $c$ and $a$. In this way $tr R$ vanishes all along the flow, as expected from string theory arguments. This result should be true for general supersymmetric theories, and for flows given by gauge or by superpotential couplings, not just for quiver gauge theories considered here.

\subsection{Superpotentials, Chiral Quivers and Seiberg Duality}
The arguments given in section (\ref{c=a}) lead naturally to the question:\\
Can any superconformal quiver be described by D3 branes at a Calabi-Yau singularity?

The support for a positive answer is provided by:
\begin{itemize}
\item The fields are in 2-index representations (leading to a natural
large $N$ limit), as is always the case for D3 brane setups.
\item Equality of the central charges $c$ and $a$. This is always true \cite{henningson} for Type IIB superstrings compactified on $AdS_5 \times X_5$, which are dual to the D3 brane theory.
\end{itemize}

In the case of $\mathcal{N} = 2$ supersymmetry it is already known \cite{vafa97}\cite{vafa01} that there is a one-to-one correspondence between superconformal quivers and geometries preserving $\mathcal{N} = 2$ supersymmetry. The latter are classified by orbifolds of $\IC^2$; the well known Mc-Kay correspondence then implies that the quiver diagrams are exactly the Dinkin diagrams of $\hat{A}, \hat{D}, \hat{E}$ algebras. On the gauge theory side it is possible to prove that these are the only possibilities \cite{vafa97}. Two key ingredients in this classification are that there are no anomalous dimensions for the elementary fields and that the form of the superpotential (we continue to use an $\mathcal{N} = 1$ language) is imposed by $\mathcal{N} = 2$ supersymmetry.

For $\mathcal{N} = 1$ supersymmetry the problem is more complicated. On the geometry side a complete classification of $3$ complex dimensional Calabi-Yau singularities is still not known. On the gauge theory side there are at least three complications: 
\begin{itemize}
\item there can be anomalous dimensions: the whole class of superconformal quivers is larger than the class given by the orbifolds of $\IC^3$, already classified in \cite{Hanany:1998sd}, where anomaluos dimensions are absent. 
\item the form of the superpotential does not descend directly from the quiver diagram. In the case of ``toric'' quivers (the ranks of the gauge groups are all equal) there is a technique, called Inverse Algorithm \cite{Feng:2000mi}, that enables to find the superpotential. One can use symmetries to restrict the possible superpotentials, as was done for ordinary symmetries in \cite{Feng:2002zw} and for hidden ``infinite coupling'' symmetries in \cite{Franco:2004}, however the general case is still elusive. 
\item there are models with matter in chiral representations of the gauge group.
\end{itemize}

\subsubsection{From superpotentials to geometry}\label{Wandgeom}
Before discussing the property of chirality, we would like to make some more comments on the superpotential of a generic $\mathcal{N} = 1$ quiver. In section (\ref{c=a}) we only considered the data that can be encoded in the quiver structure: the field content of the theory. The field content and the vanishing of the numerators of the beta functions for the gauge couplings were the only ingredients needed to show that $c = a$. However it is important to stress that this constitutes only a part of the definition of a 4D superconformal field theory: it should also be specified what is the exact form of the superpotential $\mathcal{W}$. Given the precise form of the superpotential (a polynomial of $r$-charge 2 in the chiral superfields) it is possible to calculate the moduli space of vacua of the theory, using the $\mathcal{F}$-term equations.

As is well known, parallel D3 branes do not feel forces between each other and can freely move in the transverse space, so the moduli space of the 4D gauge theory coincides with the transverse Calabi-Yau cone. More precisely the
moduli space $\mathcal{M}$ is the symmetrized product of the Calabi-Yau cone $Y$:
\beq
\mathcal{M} = Y^N/S_N\;\;,
\eeq
where $S_N$ is the symmetric group of permutations. If non-trivial fluxes are turned on, the moduli space can reduce to a subcone of $Y$: a flux for the RR $3$-form field strength can lift some directions for the motion of the D3 branes in the transverse space, since such a flux can be interpreted as giving rise to fractional D3 branes.

These properties of $\mathcal{M}$ are non trivial requirements on $\mathcal{W}$, and we are not aware of any argument showing that for any quiver it is possible to find a superpotential satisfying these requirements.

Given a superconformal quiver $\mathcal{Q}$, finding a $\mathcal{W}$ leading to a moduli space $\mathcal{M}$ which is a three (or less) complex dimensional Calabi-Yau cone should be equivalent to show that the quiver gauge theory $\mathcal{Q}$ has a Type IIB string dual: Type IIB string theory compactified on $\mathcal{M}$ should lead to a gauge theory with quiver $\mathcal{Q}$, superpotential $\mathcal{W}$ and moduli space $\mathcal{M}$, i.e. the theory we started with. This ``Reverse Geometric Engineering'' procedure has been discussed in detail by Berenstein in \cite{berenstein}.

In general finding the ``right'' superpotential, and consequently the moduli space, is not an easy task. For example, in the case of the well studied quivers arising from (complex) cones over the complex surfaces del Pezzo 7 and del Pezzo 8, the superpotential is not known, also if it is guaranteed that a superpotential which gives the correct 3 complex dimensional moduli space exists. We remark that given a quiver diagram it is not true that the ``right'' superpotential is unique. For instance in the del Pezzo $n$ surfaces (for $n \geq 5$) \cite{Wijnholt:2002qz} there is a moduli space of superpotentials, of complex dimension $2 (n - 4)$, arising from the possibility of varying the complex structures for the del Pezzo surface.

Almost all the considerations we are going to make in the next sections are just at the level of the quiver structure: quiver diagram, ranks of the gauge groups and $r$-charges of the matter fields. We will be able to find the operators entering the superpotential (they correspond to simple loops in the quiver diagram), but we will ignore the issue of which are the relative coefficients among the different monomial terms.


\subsubsection{Chiral Quivers}
$\mathcal{N} = 2$ quivers and their direct $\mathcal{N} = 1$ descendants, called ``generalized conifolds'', are non chiral models. This means that for each node the matter fields and gauge interactions are non chiral: for each bifundamental field transforming in the $( \overline{N_i}, N_j )$, there is a corresponding bifundamental field transforming in the $( \overline{N_j}, N_i )$. This doublet forms the $\mathcal{N} = 2$ hypermultiplet.

In this paper we want to study the models where the $\mathcal{N} = 1$ features are most evident, so we restrict the attention to \emph{completely chiral quivers}. With this term we mean quivers with only chiral matter: adjoints are absent and, given two nodes, arrows connecting them go only in one direction.

The matter content of a chiral quiver can be represented by a square antisymmetric matrix $\mathcal{Q}$, called \emph{quiver matrix}, of dimension equal to the total number of nodes. The element $\mathcal{Q}_{i\,j}$ is the number of arrows from node $i$ to node $j$, taken with a minus sign if the arrows go from node $j$ to node $i$.

A well studied set of completely chiral quivers are the superconformal theories arising on a stack of D3 branes placed on vanishing del Pezzo surfaces (see, for example, \cite{vafa01}, \cite{Wijnholt:2002qz, Hanany:2001py, Feng:2002zw, Feng:2002fv, Intriligator:2003wr, Franco:2003ja, Franco:2004, Franco:2003ea}).

For all the del Pezzo quivers the quiver matrix has rank two. This can be seen \cite{Hanany:2001py}\cite{Feng:2002kk} using mirror symmetry and $(p, q)$ webs techniques. The quiver matrix is then written as
\beq
\mathcal{Q}_{i\,j} = p_i q_j - q_i p_j \;,
\eeq
where $p_i$ and $q_j$ are integer numbers describing properties of the mirror picture.

The above definition of completely chiral quivers is the field theoretical analogue of the algebro-geometric notion of ``exceptional collection of sheaves'', important in the study of quivers related to cones over a del Pezzo surface (for recent work see \cite{Herzog:2003dj, Herzog:2003zc, Herzog:2004, Aspinwall:2004}). Physically the meaning is that, given two fractional branes at the singularity, there is at most one type of strings stretching between them. This tells that we have also to require that all arrows between two given nodes have the same $r$-charges. In general, verifying this property requires the knowledge of the exactly marginal superpotential. This property is always true in the case of three block models corresponding to del Pezzo surfaces: del Pezzo $0$, $\IF_0$ and del Pezzo $n$ with $3 \leq n \leq 8$.

\subsubsection{Duality Trees}\label{dualitytrees}
Chiral quivers are also the natural arena for the study of Seiberg Duality \cite{Seiberg:1994pq}. In general, if a gauge group $U(N)$ is Seiberg Dualised, the resulting quiver is different from the initial one\footnote{In the ``generalized conifold'' Seiberg Duality doesn't change the quiver diagram and the $r$-charges (the Duality Tree has just one node). However this doesn't mean that Seiberg Duality is trivial in these theories, since it changes the values of the couplings \cite{vafa01}. For recent work see \cite{warner04}.}. Applying again Seiberg Duality to another node, a new different quiver is obtained, and so on.

Also if the quiver matrix changes, the rank of the quiver matrix is invariant under Seiberg Duality.

An equivalence class of Seiberg dual quivers is in general composed by infinite different quivers, and can be organized in a ``Duality Tree'' \cite{Franco:2003ja} \cite{Franco:2003ea}. A Duality Tree is a graph where nodes are represented by theories, and two nodes are connected by an edge if there is a Seiberg Duality leading from one theory to the other. We remark that in general there can be loops in this tree. These loops give the possibility of engineering Duality Cascades of the Klebanov-Strassler type \cite{klebanovstrassler}, \cite{Franco:2003ja}, \cite{Franco:2004jz}.

A simple general property of Duality Trees that will be important for our classification results is the existence of the ``roots of the tree'', or of the ``minimal models''. We define the roots to be those models of the Duality Tree that (locally) \emph{minimize the sum of the ranks of the gauge groups}. That is models for which Seiberg Duality on any node increases the rank of that node.

\begin{figure}[h]
  \epsfxsize = 4 cm
  \centerline{\epsfbox{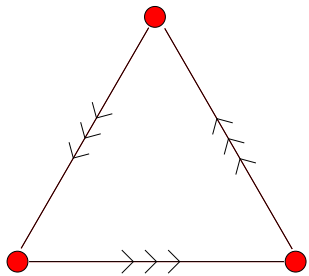}}
  \caption{One of the quivers associated to $dP_0 \equiv \mathbb{CP}^2$. This is the only root of its associated Duality Tree, also called the Markov Tree.}
  \label{quiver_dP0}
\end{figure}

From the definition, since the sum of the ranks of the gauge groups is a positive integer, it is obvious that for every Duality Tree there is at list one root. However we would like to describe these minimal models more explicitly. In order to have a better understanding we define the ``relative number of flavors'' $\nf$  of a node in a chiral quiver to be
\beq\label{nfdef}
\nf_i = \frac{N_f}{N_c} = \frac{1}{N_c} \sum_{j} \frac{1}{2} |\mathcal{Q}_{i,\,j}|\, N_j \;,
\eeq
i.e. the ratio between the total amount of matter charged under the gauge group and the rank of the group.\\
For example in supersymmetric orbifolds of $\IC^3$ (classified by discrete subgroups of $SU(3)$, \cite{Hanany:1998sd}) all the nodes have $\nf = 3$. One of the simplest is the del Pezzo $0$ quiver of figure \ref{quiver_dP0}. In fact this is the well known orbifold $\IC^3/\IZ_3$.

Seiberg dualizing the $i^{th}$ node of a quiver the rank of the corresponding gauge group changes as
$$ N_c \rightarrow N_f - N_c\;,$$
so (since $N_i = x_i \, N$)
\beq
x_i \rightarrow x_i \, (\nf_i - 1)\;.
\eeq
From this formula it follows that $x_i$ decreases if $\nf_i < 2$, otherwise it increases. Thus, given a quiver in  the Duality Tree, if there is a node with $\nf < 2$, Seiberg dualizing this node reduces the sum of the ranks of the gauge groups. In other words we are moving downward in the Duality Tree. This process can be repeated until the reach of a quiver such that $\nf_i \geq 2$ for all $i$. At this point it is not possible to decrease the ranks of the groups anymore: this theory is a root (or minimal model) of the Duality Tree.

The conclusion is that minimal models are exactly the models with $\nf \geq 2$ for each gauge group.

It is important to note that the roots are in general not unique. For instance there could be nodes with $\nf = 2$, dualizing one of them doesn't change the ranks of the groups but can change the quiver diagram and the total number of bifundamental fields. We will encounter other examples of this fact in the next section, when we will see that there are different ``3-block'' minimal quivers in the same Duality Tree.

The existence of roots is useful in the classification of quivers: all the quivers are obtained by applying successive Seiberg Dualities to quivers satisfying the minimality condition, thus it is enough to classify quivers with $\nf \geq 2$ for each gauge group.

We would like to remark that these simple arguments do not require the chirality of the quiver, also if for simplicity we only discussed this case. Of course for non-chiral quivers the definition of $\nf$ is different, for instance also adjoint fields contribute.

\section{Three-block chiral quivers}\label{threeblock}
In this paper we want to start the classification of chiral quivers, taking in consideration models with a quiver matrix of rank equal to two, the minimal possible rank. We have not been able to classify all rank two chiral quivers. The main problem is that it is difficult to perform a general analysis of quivers with five or more nodes: it is not clear which loops enter in the superpotential, and it is not completely clear if an arbitrary number of Seiberg Dualities leaves the quiver chiral. Recently there has been work devoted to understanding the precise conditions ensuring that the Seiberg dual of a chiral quiver is chiral (this leads to the notion of ``strongly exceptional collection'', see \cite{Herzog:2003zc} \cite{Herzog:2004} \cite{Aspinwall:2004}).

A way to circumvent these problems is to consider quivers that can be organized in three or four blocks.

The nodes of a quiver can be grouped in ``blocks'', if
\begin{itemize}
\item each node in a given block has the same rank;
\item there are no arrows between nodes of the same block;
\item the number and the direction of arrows between two nodes depend
only on the blocks where the nodes are.
\end{itemize}

Equivalently the quiver matrix and the vector of the ranks of the gauge groups satisfy a block structure.

Instead of analyzing from the beginning the four-block quivers (which includes as a particular case the three-block), we first discuss the three-block quivers. The reason is that in this simpler case it is possible to discuss some interesting features (unitarity and asymptotic freedom) more easily. Moreover we arrive at the definition of the procedure of shrinking, to be discussed in the next section. Surprisingly it turns out that using this procedure it is possible to obtain all the four-block superconformal quivers starting from the known three block del Pezzo quivers. It also turns out that it is not possible to obtain the models associated to del Pezzo $1$ and del Pezzo $2$. This is check of our procedure, since these two models have irrational $r$-charges and the shrinking procedure does not change the $r$-charges.

\subsection{General structure of three-block quivers}
In this section we analyze in detail the case of chiral quivers that can be organized in a three-block structure, i.e. in their Duality Tree there is sub-tree composed by three-block quivers. We study the constraints on the matter content of the quiver arising by anomaly cancellation and by conformal invariance. This leads to a diophantine equation for the entries of the quiver matrix. This equation depends on the dimensions of the three blocks and classifies all possible models. In the simplest case of all three blocks of dimension one (i.e. a 3 nodes quiver) it reduces to the well known Markov equation and the corresponding Duality Tree is called the Markov Tree. We have been able to prove that, modulo Seiberg Duality, exactly 10 different theories admit a three-block structure. These are the 8 quivers arising from the del Pezzo surfaces $\IC\IP^2$, $\IF_0$, $dP_n$ ($3 \leq n \leq 8$) and just two other models. These two models can be obtained by ``shrinking" $4$ nodes of the del Pezzo 5 and the del Pezzo 7 quivers.

The diophantine equations related to del Pezzo surfaces have already been studied by mathematicians \cite{Karpov:1} in work on three-block exceptional collections. Here however we will not restrict ourselves to del Pezzo surfaces. Our new result is that there are only two other possible diophantine equations. We leave the detailed proof of the unicity of this 10 models to appendix \ref{classification}.

\vspace{0.4 cm}

\begin{figure}[h]
  \epsfxsize = 9 cm
  \centerline{\epsfbox{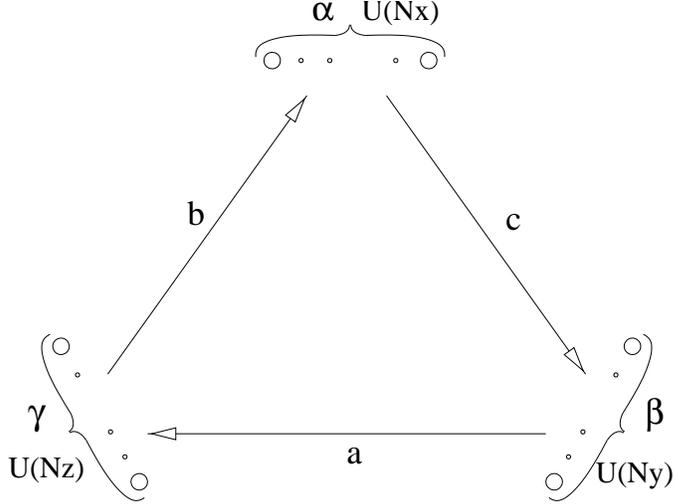}}
  \caption{General 3-block chiral quiver diagram.
         $\alpha, \beta, \gamma$ are the number of nodes in each block.
         $a, b, c$ are the number of bifundamental connecting the nodes.}
  \label{quiver_3block}
\end{figure}

For three blocks the structure of the quiver diagram is always a triangle, since for each block there is exactly one incoming arrow and one outgoing arrow. There is thus just one way of drawing the quiver, as in figure \ref{quiver_3block}. 
In this Figure $\alpha, \beta, \gamma$ are the number of nodes in each block, $a$, $b$ and $c$ are the number of bifundamentals connecting the nodes. The gauge groups are $U(N\,x)$, $U(N\,y)$, $U(N\,z)$. The nine integer numbers $( \alpha, \beta, \gamma; \, a, b, c; \, x, y, z )$, together with the parameter $N$, define completely the field content of the theory.

The quiver matrix is a square antisymmetric block-matrix of dimension $(\ag + \bg + \cg$), with blocks of dimension $\alpha \times \alpha$, $\alpha \times \beta$ etc.

The first physical condition to be imposed on the theory is that for each gauge group the ABJ gauge anomalies vanish:
\beq\label{ABJconstraint}
\sum_j \mathcal{Q}_{i j} ( N x_j ) = 0 \;\;\;\; \text{for each}\;i.
\eeq
The kernel of $\mathcal{Q}$ has dimension \mbox{$( dim[\mathcal{Q}] - 2 )$}, but we have to search this kernel for vectors satisfying the ``three-block'' condition:
$$\big( \underbrace{x, \ldots, x}_{\ag}, \, \underbrace{y, \ldots, y}_{\bg}, \, \underbrace{z, \ldots, z}_{\cg} \big) \,.$$
It is convenient to consider a reduced antisymmetric quiver matrix:
\beq\label{reducedmatrix} 
q = \left(
\begin{array}{ccc}
    0   &  c   &  -b   \\
   - c  &  0   &   a   \\
    b   & - a  &   0   \\
\end{array}
\right) \eeq Canceling the gauge anomalies means that the 3-vector $(\ag x, \bg y, \cg z)$ has to be in the kernel of $q$. In the 3-blocks chiral quivers there is always exactly one such vector, modulo rescaling, since a $3 \times 3$ antisymmetric matrix has always rank $2$ (excluding the trivial case \mbox{$q = 0$}).

The kernel of $q$ is generated by a vector proportional to $(a, b, c)$, so $( x, y, z )$ have to satisfy:
\beq\label{anomaly_3block}
(\ag x, \bg y, \cg z) \propto (a, b, c)\;.
\eeq

At this point, for any choice of $\ag, \bg, \cg$ and of the $3 \times 3$ antisymmetric matrix $q$ there are well defined gauge theories (if the vector $(\ag x, \bg y, \cg y)$ lies in the kernel of $q$), however it is not guaranteed that the space of couplings contains an interacting conformal fixed point.

In order to achieve this last requirement, which is quite restrictive, the beta functions of the gauge couplings and of the superpotential couplings have to vanish. Once the field content of a supersymmetric theory is given, the exact beta functions are completely controlled by the $r$-charges of the matter fields.

We call the $r$-charges of the three set of bifundamental chiral superfields $r_a, r_b, r_c$. The beta functions of the gauge couplings, since we require an \emph{interacting} RG fixed point, will vanish exactly when their numerators vanish (see formulae (\ref{NSVZa}) and (\ref{NSVZb})):
\bea\label{NSVZconstraint3a}
N x + \frac{1}{2} \bg N y \, c (r_c - 1) + \frac{1}{2} \cg N z \, b (r_b - 1) &=& 0\nonumber\\
N y + \frac{1}{2} \cg N z \, a (r_a - 1) + \frac{1}{2} \ag N x \, c (r_c - 1) &=& 0\\
N z + \frac{1}{2} \ag N x \, b (r_b - 1) + \frac{1}{2} \bg N y \, a (r_a - 1) &=& 0\nonumber\;.
\eea
Using (\ref{anomaly_3block}) these equations can be written in terms of the quiver numbers $(a, b, c)$:
\bea\label{NSVZconstraint3b} 
\frac{a}{\ag} + \frac{1}{2}  b \, c (r_c - 1) + \frac{1}{2} c \, b (r_b - 1) &=& 0\nonumber\\
\frac{b}{\bg} + \frac{1}{2}  c \, a (r_a - 1) + \frac{1}{2} a \, c (r_c - 1) &=& 0\\
\frac{c}{\cg} + \frac{1}{2}  a \, b (r_b - 1) + \frac{1}{2} b \, a (r_a - 1) &=& 0\nonumber\;\,.
\eea
There is also a relation coming from the vanishing of the beta functions of the superpotential couplings. From the quiver diagram in (\fref{quiver_3block}) it is clear that all gauge invariant operators candidate to appear in the superpotential are cubic\footnote{There could be a priori the possibility of taking a superpotential sextic (or of higher order) in the fields. In these cases it can be shown, using exactly the same methods of appendix \ref{classification}, that the corresponding diophantine equations have no solutions.} and can be schematically written as \beq \mathcal{W} = X_{\alpha \beta}X_{\beta \gamma}X_{\gamma \alpha}\;. \eeq We would like to stress that this does not give enough information to construct the complete superpotential and then the moduli space of vacua. This is for two reasons: first there can be arrows with multiplicity (some entries of the quiver matrix are greater than $1$), second in the blocks there can be more than one node. So there are many gauge invariant operators corresponding to $X_{\alpha \beta}X_{\beta \gamma}X_{\gamma \alpha}$, and it should be specified which are the relative coefficients of these different operators in the superpotential.

A chiral operator is marginal at the interacting fixed point if its total $r$-charge is $2$. So the $r$-charges have to satisfy another relation: 
\beq\label{Wconstraint3b} 
r_a + r_b + r_c = 2\,. 
\eeq 
In total there are $4$ linear equations in the $3$ variables $r_a, r_b, r_c$. In order that a solution exists the integers $(a, b, c)$ have to satisfy a constraint. 
This constraint can be readily obtained by considering, like in section (\ref{c=a}), the sum of the three beta functions (\ref{NSVZconstraint3b}) (weighted respectively by the numbers $a$, $b$, $c$): 
\beq 
\frac{a^2}{\ag} + \frac{b^2}{\bg} + \frac{c^2}{\cg} + a b c (r_a + r_b + r_c - 3) = 0 \,.
\eeq 
The superpotential thus imposes (\ref{Wconstraint3b}) the following diophantine equation: 
\beq\label{dioph3blocks}
\frac{a^2}{\ag} + \frac{b^2}{\bg} + \frac{c^2}{\cg} = a b c \,.
\eeq

This equation is a generalization of the well known Markov Equation ($\ag = \bg = \cg = 1$); it automatically encodes all possible Seiberg Dualities, which act keeping fixed $( \ag, \bg, \cg )$ and changing one of $( a, b, c )$, depending on which block is dualized. The reason is as follows: (\ref{dioph3blocks}) is quadratic in each of the three variables $( a, b, c )$, solving it in terms of one of the three variables (for instance $a$) gives of course $2$ solutions:
\beq\label{soldiophdegree2}
a = \frac{\ag b c \pm \sqrt{(\ag b c)^2 - 4 \ag \, \left(\frac{b^2}{\bg} + \frac{c^2}{\cg}\right)}}{2}\;.
\eeq
Now suppose we know a particular integer solution $( a_0, b_0, c_0 )$ of (\ref{dioph3blocks}), corresponding to the superconformal quiver $\mathcal{Q}$. The triple $( a_0, b_0, c_0 )$ clearly satisfies (\ref{soldiophdegree2}) as well, for a specific choiche of the sign in the r.h.s.. Since $a_0$ is integer, the expression inside the square root is a perfect square, so choosing the other sign gives another triple $( \tilde{a}_0, b_0, c_0 )$. This will be another integer solution of (\ref{dioph3blocks}), corresponding to the superconformal quiver obtained by $\mathcal{Q}$ after Seiberg Duality of the block with $\ag$ nodes. To be precise we should also note that, since $a_0$ is integer, if $\ag b c$ is even (odd) also $\sqrt{\ldots \, }$ is even (odd), so the numerator in (\ref{soldiophdegree2}) is even for both choices of the sign.

It is worth noting that (\ref{dioph3blocks}) doesn't only tell that there is a Duality Tree, but also ensures that all the models in the Duality Tree are completely chiral. Indeed the diophantine equation could also be derived by requiring chirality for the Seiberg dualized quiver \cite{Herzog:2003zc} or by studying ``Partial Seiberg Dualities'' (Picard-Lefschetz transformations) \cite{Feng:2002kk}. We decided to use just the superconformal conditions because in this way the chirality of the full Duality Tree is a consequence. Since in the $4$-block case we will follow the same procedure, also in that case it will be ensured that the full Duality Trees of the models we will find are chiral. 

It is however possible to understand directly the chirality of the full Duality Tree. The reason is that the magnetic mesons are represented by arrows in the opposite direction with respect to the original bifundamental fields (the ones in the edge opposite to the dualized block). Since we imposed the presence of the cubic superpotential, the $r$-charge of the mesons is exactly $2$ minus the $r$-charge of the oppositly directed field, so all of these last fields can be integrated out using the quadratic term in the superpotential that couples the magnetic mesons to these last fields. In other words all double arrows formed in a Seiberg Duality of a block are marginal, in the sense that they have total $r$-charge $2$, so they can be integrated out with a marginal quadratic term in the superpotential.

\vspace{0.2 cm}

We have been able to classify all possible triples ($\ag, \bg, \cg$) leading to an integer solution for (\ref{dioph3blocks}). The detailed analysis is reported in appendix \ref{classification}. The result is that there are only 16 such triples and that the total number of gauge groups $\ag + \bg + \cg$ is at most $11$. Before giving the list of these models, we would like to discuss some general properties of three block quivers that can be easily obtained from equation (\ref{dioph3blocks}). We will pause on some details because these properties also hold for the four-block chiral quivers, under certain assumptions on the superpotential. We think that these are general features of every physical superconformal quiver.

\subsection{$R$-charges and unitarity}\label{unitarity}
Keeping track of the diophantine equation (\ref{dioph3blocks}),
the solution of the system of the beta function equations
(\ref{NSVZconstraint3b}) is:
\beq\label{r-charges}
\left\{
\begin{array}{c}
r_a  = \frac{2 a}{\ag b c}\\
r_b  = \frac{2 b}{\bg a c}\\
r_c  = \frac{2 c}{\cg a b}\,.
\end{array}
\right.
\eeq
A first non trivial result is that all the $r$-charges, even if they change under Seiberg Duality,  are always positive. The consequence is that all gauge invariant chiral operators satisfy the unitarity bound $R \geq \frac{2}{3}$. In the case of three block this is always true, but with four blocks it is possible to construct candidate superconformal quivers such that some $r$-charges are negative \cite{Herzog:2003zc}\cite{Feng:2002kk}. With the fields of negative $r$-charge it possible to construct scalar gauge invariant baryonic operators of negative dimension, and unitarity of the gauge theory is violated. Such operators correspond holographically to states violating the Breitenloner-Friedmann bound, i.e. tachyons on the Anti-de-Sitter space. We will see that these models, that can be called ``tachyonic quivers'', do not satisfy the corresponding four-block diophantine equation. It is possible to construct examples of tachyonic quivers applying ``Partial Seiberg Dualities'' \cite{Feng:2002kk} to physical quivers. In \cite{hanany04} it has been argued that these models ``decay'' in a new different quiver, with a larger number of nodes and a higher rank for the quiver matrix. In this paper however, imposing the full superconformal constraints, we rule out all possible tachionic quivers.

\vspace{0.3 cm}

Another check of the unitarity of the three-block gauge theory is the positivity of the gravitational central charge\footnote{Since $c$ is related to the two point function of the energy-momentum tensor unitarity implies $c > 0$. One of the statements of the conjectured $a$-theorem is that also $a$ is positive for unitary theories. However in this paper all the models satisfy $c = a$, so positivity of $a$ follows from positivity of $c$.} $c$. Since $tr R = 0$, equations (\ref{AFGJ}) imply that the central charges are proportional to $tr R^3$. Apart from a factor of $N^2$ from (\ref{traceR}) we have:
\beq\label{trR3}
tr R^3 = \ag  x^2 + \bg  y^2 +
\cg  z^2 + \ag \bg x y c (r_c - 1)^3 + \ag \cg  x z b (r_b - 1)^3 + \bg \cg  y z a (r_a - 1)^3\;\,.
\eeq
At this point it is necessary to specify the constant of proportionality (\ref{anomaly_3block}) between the ranks $(x, y, z)$ and $(a/\ag, b/\bg, c/\cg )$. We choose this costant to be $\sqrt{( \ag \bg \cg ) / K^2}$, leading to:
\beq\label{Ksq} \left\{
\begin{array}{l}
K^2 x^2 = a^2 \frac{\bg \cg}{\ag}\\
K^2 y^2 = b^2 \frac{\ag \cg}{\bg}\\
K^2 z^2 = c^2 \frac{\ag \bg}{\cg}\;\,.
\end{array}
\right. \eeq $K^2$ is a priori a rational number defined by the property that the numbers $(x, y, z)$ are integer and have no common factors.

The reason for this apparently complicated choice is that it turns out that $K^2$ is always an integer satisfying a simple relation with the total number of nodes of the quiver. The name $K^2$ (it is not a perfect square in general) has been chosen in order to make contact with literature on del Pezzo quivers, where $K^2$ is the square of the canonical class of the del Pezzo surface.

Inserting (\ref{r-charges}) and (\ref{Ksq}) in the formula (\ref{trR3}) gives an expression for $tr R^3$ in terms of $a, b, c, \ag, \bg, \cg$ and $K^2$. Keeping carefully track of the diophantine equation (\ref{dioph3blocks}) it is possible to show that this expression simplifies to:
\beq
tr R^3 = \frac{24 N^2}{K^2} 
\eeq 
\beq \label{canda}
c = a = \frac{9}{32} tr R^3 = \frac{27 N^2}{4\,K^2}\;.
\eeq

Among other things this result shows that $K^2$ is invariant under Seiberg Duality, since the 't Hooft anomalies are invariant as well \cite{Seiberg:1994pq}. The fact that the central charges depend only on $K^2$ also holds for all the four-block quivers, to be discussed in section \ref{fourblock}. We expect this to be valid for every physical superconformal quiver.

If the quiver comes from  D3 branes probing del Pezzo singularities this result is well-known; $K^2$ is proportional to the volume of the del Pezzo surface. AdS/CFT correspondence thus tells us that it has to be inversely proportional to the central charges. We see here that this result is general. The difference between del Pezzo and ``new" quivers is the relation between $K^2$ and the total number of nodes:
\beq \label{KSQ-nodes}
\left\{
\begin{array}{lc}
K^2  = 12 - (\ag + \bg +\cg)  &  \text{for del Pezzo quivers}\\
K^2  = 9  - (\ag + \bg +\cg)  &  \text{for the ``new'' quivers}
\end{array}
\right. \eeq
We will give an explanation of this result in the next section, while discussing the procedure of ``shrinking''. We remark that this relation is a consequence of our classification, and we arrived at it looking at all the 16 possible cases. In particular it cannot be used to prove a priori that for any 3-block chiral quiver the number of nodes is at most $11$.

\subsection{Asymptotic Freedom for quivers}
Another general fact we would like to discuss is the following ``asymptotic freedom'' property:

\begin{quote}
\emph{it is always possible to flow to the superconformal interacting
      fixed points starting from the free ultraviolet fixed point.}
\end{quote}

This property is true also for the four-block models we will find in section \ref{fourblock}. It is interesting to note that the ``tachyonic quivers'' usually do not satisfy this requirement (and they also do not satisfy the four-block diophantine equation).

In a supersymmetric theory, a $SU(N_c)$ gauge group is asymptotically free if the total amount of matter $N_f$ satisfies $N_f \leq 3 N_c$. We define a quiver to be  asymptotically free if at least one of the nodes has number of flavors $\nf \leq 3$ ($\nf$ has been defined in section \ref{dualitytrees}). In this case near the free points there are directions with negative beta function for at least one coupling, so it is possible to flow away from the trivial fixed point and arrive at the interacting conformal theory, which is thus an IR fixed point.

If all the gauge groups have $\nf = 3$, each gauge coupling is marginally irrelevant, so one would suspect that it is impossible to flow away from the free fixed point. However in this case there are always exactly marginal directions: a manifold of fixed points passing through the free theory. The reason is that it is possible to find a combination of the gauge and superpotential couplings such that the corresponding beta function is exactly zero. This can be shown using Leigh-Strassler type arguments \cite{leighstrassler}. We remark that the free ultraviolet theory, if not lying on the fixed point manifold, does not satisfy $c = a$.

\vspace{0.3 cm}

It is easy to show that for all three-block chiral quivers it is possible to flow away from the free theory. From the quiver diagram, using (\ref{Ksq}), it is possible to find the relative number of flavors $\nf$ of the three different types of nodes:
\beq
\left\{
\begin{array}{c}
\nf_x  = c \frac{y}{x} = \ag \frac{b c}{a}\\
\nf_y  = \bg \frac{a c}{b}\\
\nf_z  = \cg \frac{a b}{c}\,.
\end{array}
\right. 
\eeq 
Comparing with (\ref{r-charges}) we see that the number of flavors of a node is proportional to the inverse of the $r$-charge of the bifundamental ``opposite'' to the node:
\beq\label{flavours} \left\{
\begin{array}{c}
\nf_x  = \frac{2}{r_a}\\
\nf_y  = \frac{2}{r_b}\\
\nf_z  = \frac{2}{r_c}\,.
\end{array}
\right.
\eeq
So (\ref{Wconstraint3b}) can be rewritten as:
\beq
\frac{1}{\nf_x} + \frac{1}{\nf_y} + \frac{1}{\nf_z} = 1\,.
\eeq
From this equation it is clear that at least one of the $\nf$ has to be less than $3$. If the smallest $\nf$ is exactly $3$ than all three $\nf$ are equal to $3$ and the free theory lies on the superconformal manifold. This concludes our simple proof of the asymptotic freedom for quivers possessing an interacting superconformal point in the space of the couplings.

We would like to remark that this result is not obvious a priori. There is a logical possibility of a situation like the following: a superconformal interacting fixed points in the ultraviolet and a free theory in the infrared. We didn't find any compelling evidence against this situation, we can just note that it would contrast with the expectation of the irreversibility of the renormalization group flow:  starting from the IR theory it should be impossible to reconstruct backward the strongly coupled UV theory, since the reverse process (flow to the IR) kills degrees of freedom and information.

\vspace{1 cm}

In the simple case of three block quiver we have thus been able to show the connections between the following expected physical properties of superconformal quivers:
\begin{itemize}
\item tachyonic operators are absent (all the $r$-charges of the bifundamental fields are strictly positive) and the trace anomalies satisfy the required positivity property.
\item Asymptotic Freedom: it is always possible to flow to the superconformal manifold starting from the free theory.
\end{itemize}
We will see that these facts are strictly related also in the four-block case, so we conjecture that they are general physical requirement any superconformal quiver satisfies. A crucial ingredient needed to arrive to the previous properties is the fact that all the bifundamental fields enter in the superpotential: this gives the diophantine equation constraint, which in turn implies the three listed properties.

\vspace{1 cm}

The results of appendix \ref{classification} show that there are just $16$ classes of solutions to the diophantine equations, leading to a total of $16$ duality trees. These represent all the chiral quivers that can be Seiberg Dualised to a three-block model. In fact not all these theories are different. For instance the gauge theories corresponding to the surface del Pezzo $6$ admits two different minimal three-block structures: one with $(\ag, \bg, \cg ) = ( 3, 3, 3 )$ and one with $(\ag, \bg, \cg ) = ( 1, 2, 6 )$. One class can be obtained from the other by successive Seiberg dualities, breaking the three-block structure in the intermediate steps. For del Pezzo $7$ there are $3$ different $3$-block minimal models and for del Pezzo $8$ there are $4$.

In total thus there are 10 different classes of quivers which admit a three-block structure. 8 of them are the well known del Pezzo quivers. Surprisingly enough there are just two ``new'' quivers. The 10 different classes are summarized in a table below (we also include, without figure, the other $6$ $3$-block models). In the next section we will find a way to construct other rank $2$ chiral quivers starting from the del Pezzo quivers.

\newpage
\begin{center}
\begin{table}[h!]
\begin{center}
\begin{tabular}{|c|ccc|ccc|ccc|c|c|}
\hline
  Quiver                                    &$\ag$&$\bg$&$\cg$&  a & b & c & x & y & z   &$K^2$ & surface\\
\hline\hline
\tsg{0.5cm}{1.5cm}\rb{-0.5cm}{\ig[height=1.4 cm]{delpezzo0.eps}}&1&1&1 &3&3&3 & 1&1&1&  9   &  $dP_0$\\
\hline
\tsg{0.5cm}{1.5cm}\rb{-0.5cm}{\ig[height=1.4 cm]{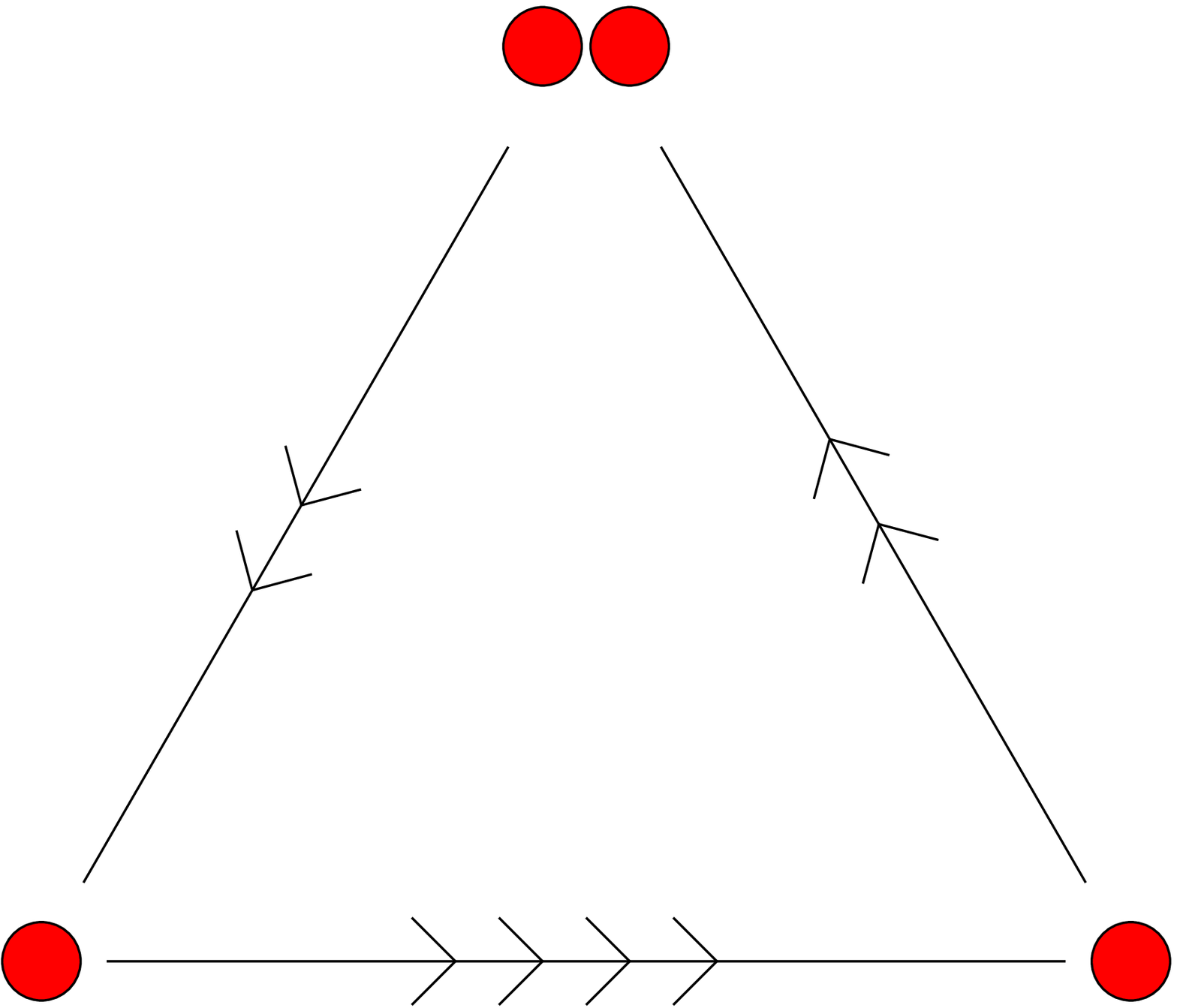}} & 1&1&2  &  2&2&4 &  1&1&1   &   8  &  $F_0$\\
\hline
\tsg{0.5cm}{1.5cm}\rb{-0.5cm}{\ig[height=1.4 cm]{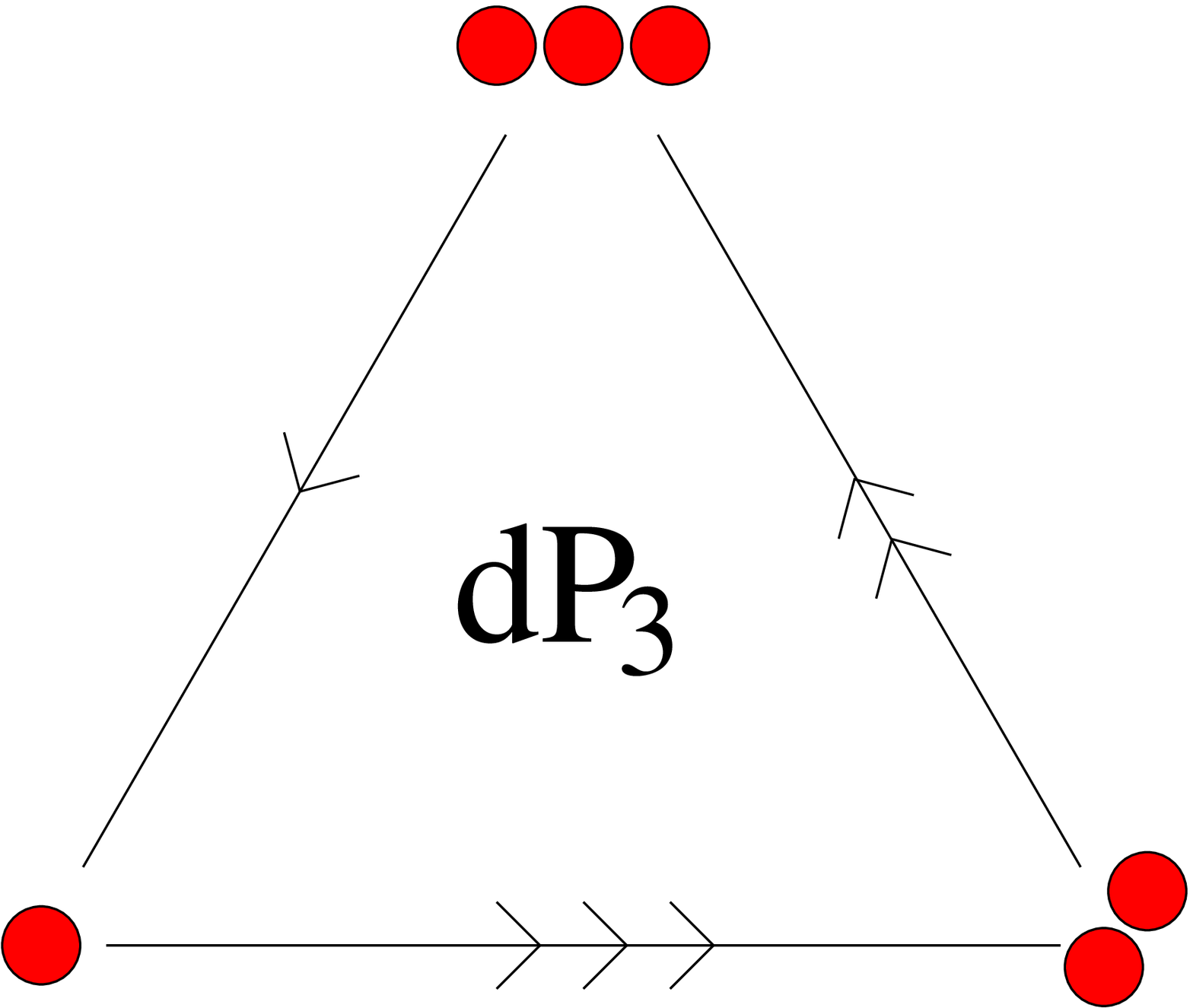}} & 1&2&3   & 1&2&3 & 1&1&1 & 6 & $dP_3$\\
\hline
\tsg{0.5cm}{1.5cm}\rb{-0.5cm}{\ig[height=1.4 cm]{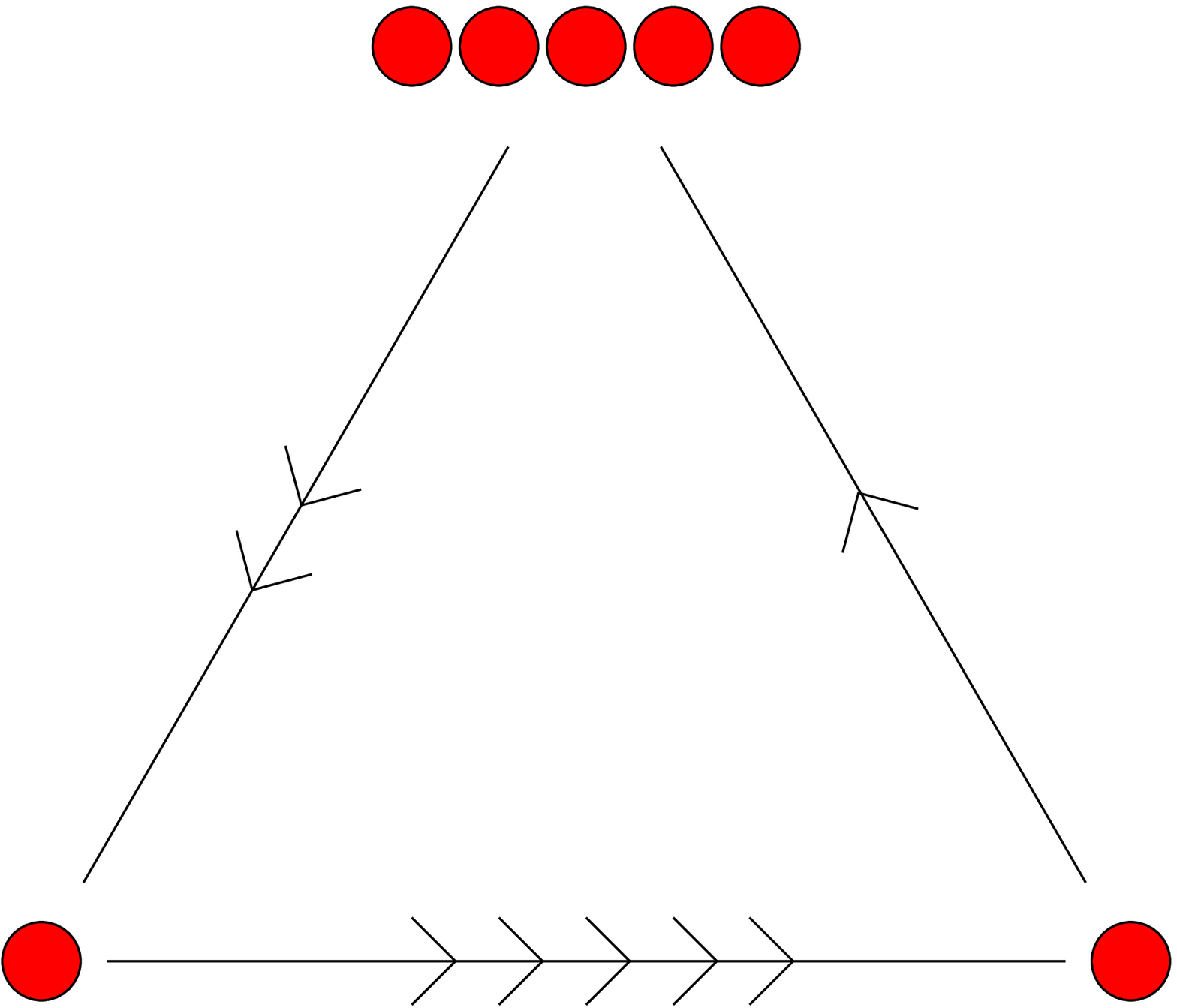}}&1&1&5  & 1&2&5 &  1&2&1   &5  &  $dP_4$\\
\hline
\tsg{0.5cm}{1.5cm}\rb{-0.5cm}{\ig[height=1.4 cm]{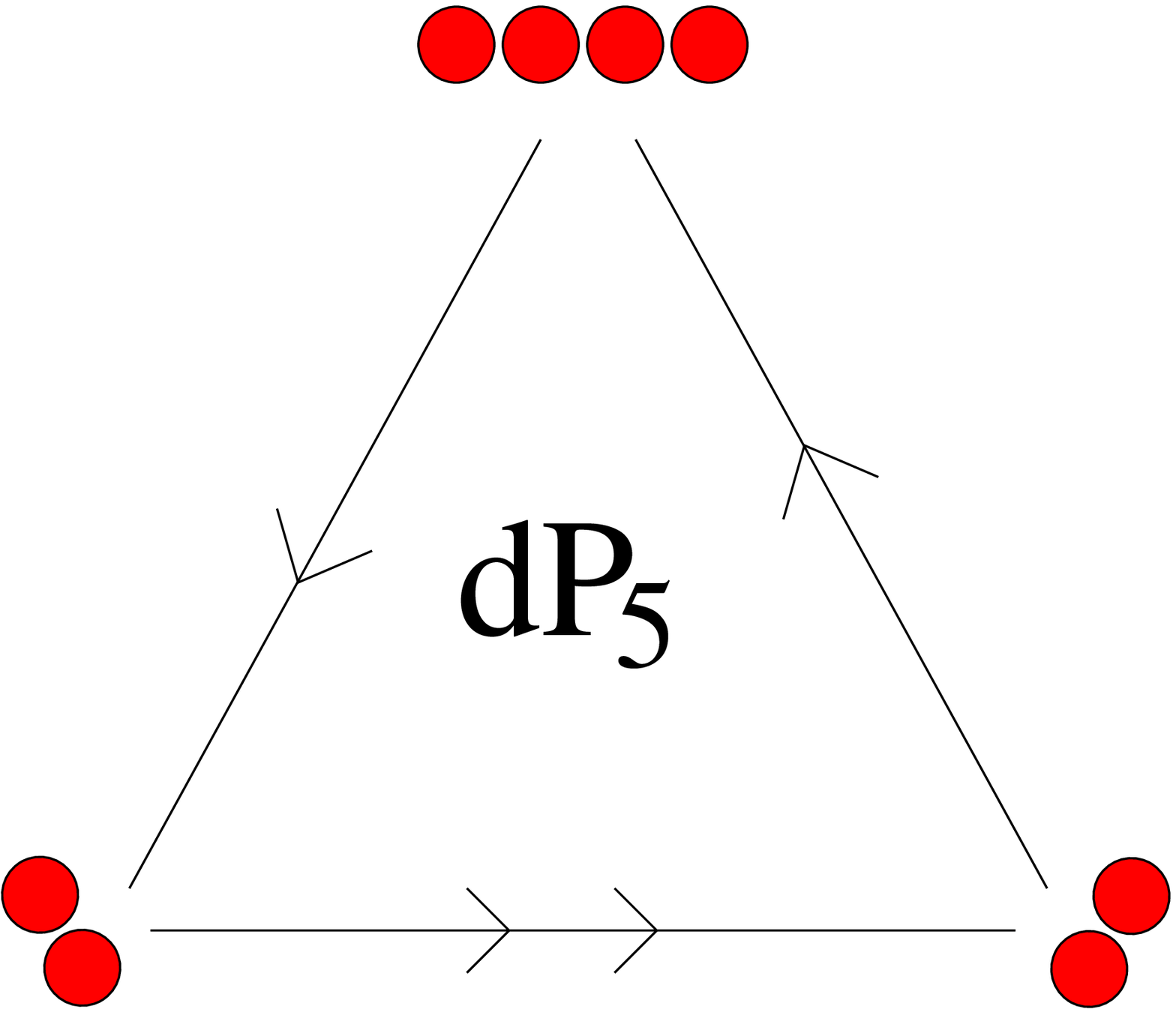}}&2&2&4 &1&1&2 & 1&1&1  & 4  &  $dP_5$\\
\hline
\tsg{0.5cm}{1.5cm}\rb{-0.5cm}{\ig[height=1.4 cm]{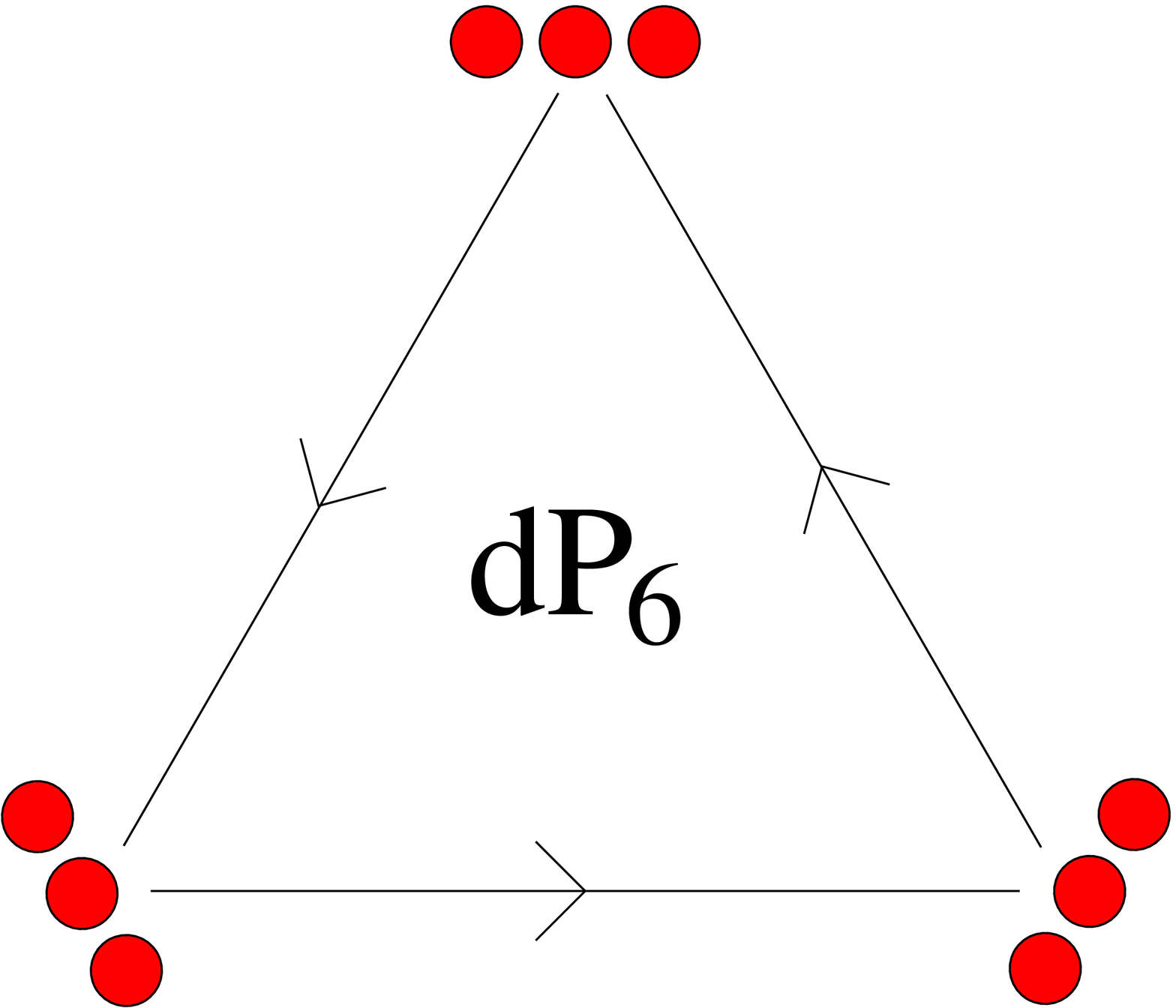}}&3&3&3 & 1&1&1 &  1&1&1 & 3 & $dP_6\;I$\\
\hline
&2&1&6 &1&1&3 & 1&2&1  &3& $dP_6\;II$\\
\hline
\tsg{0.5cm}{1.5cm}\rb{-0.5cm}{\ig[height=1.4 cm]{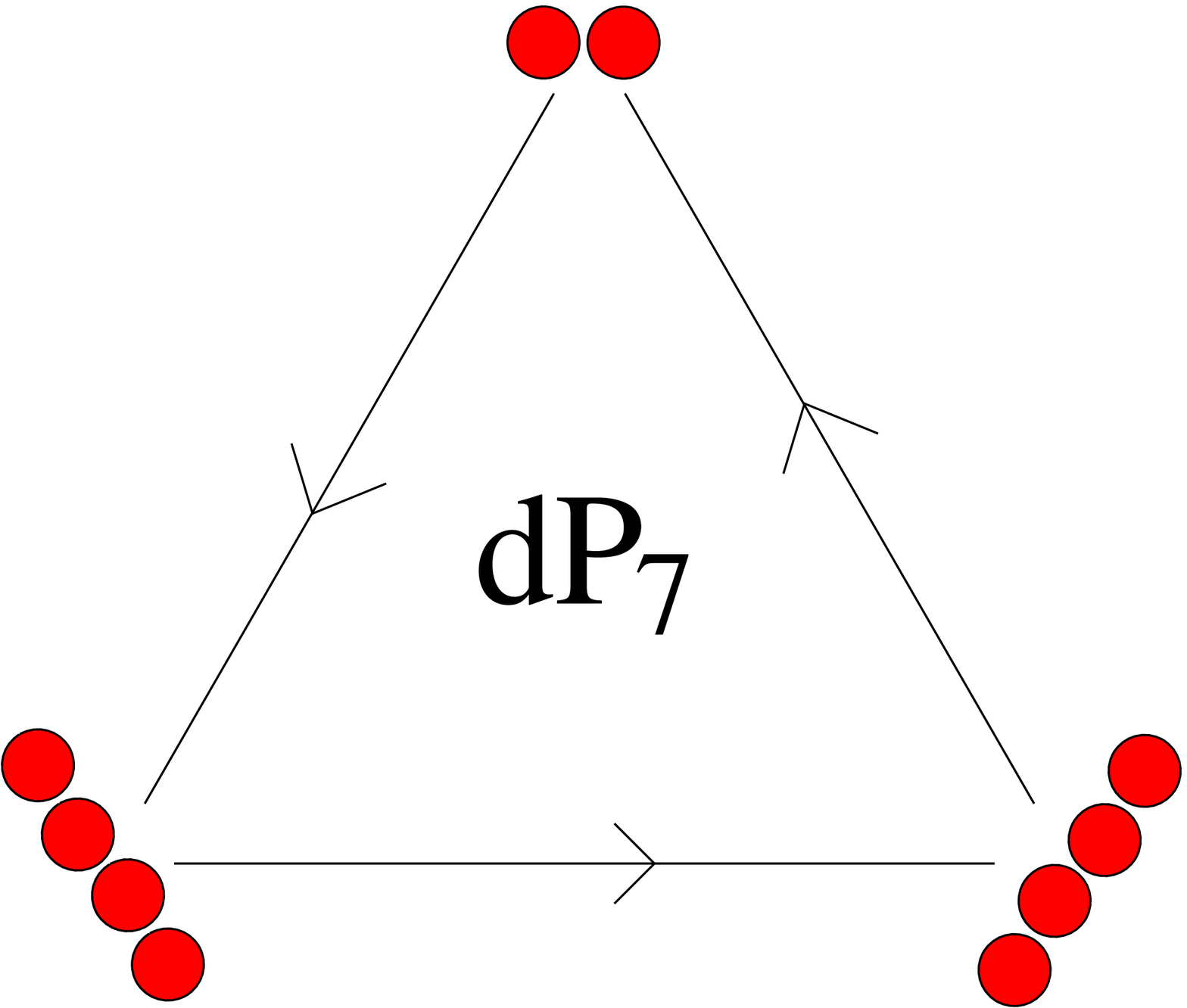}}&4&4&2 & 1&1&1& 1&1&2 &2  &  $dP_7\;I$\\
\hline
& 1 & 1 & 8    &  1 & 1 & 4 &    2 & 2 & 1    & 2  &  $dP_7\;II$ \\
\hline
& 3 & 1 & 6    &  1 & 1 & 2 &    1 & 3 & 1    & 2  &  $dP_7\;III$ \\
\hline
\tsg{0.5cm}{1.5cm}\rb{-0.5cm}{\ig[height=1.4 cm]{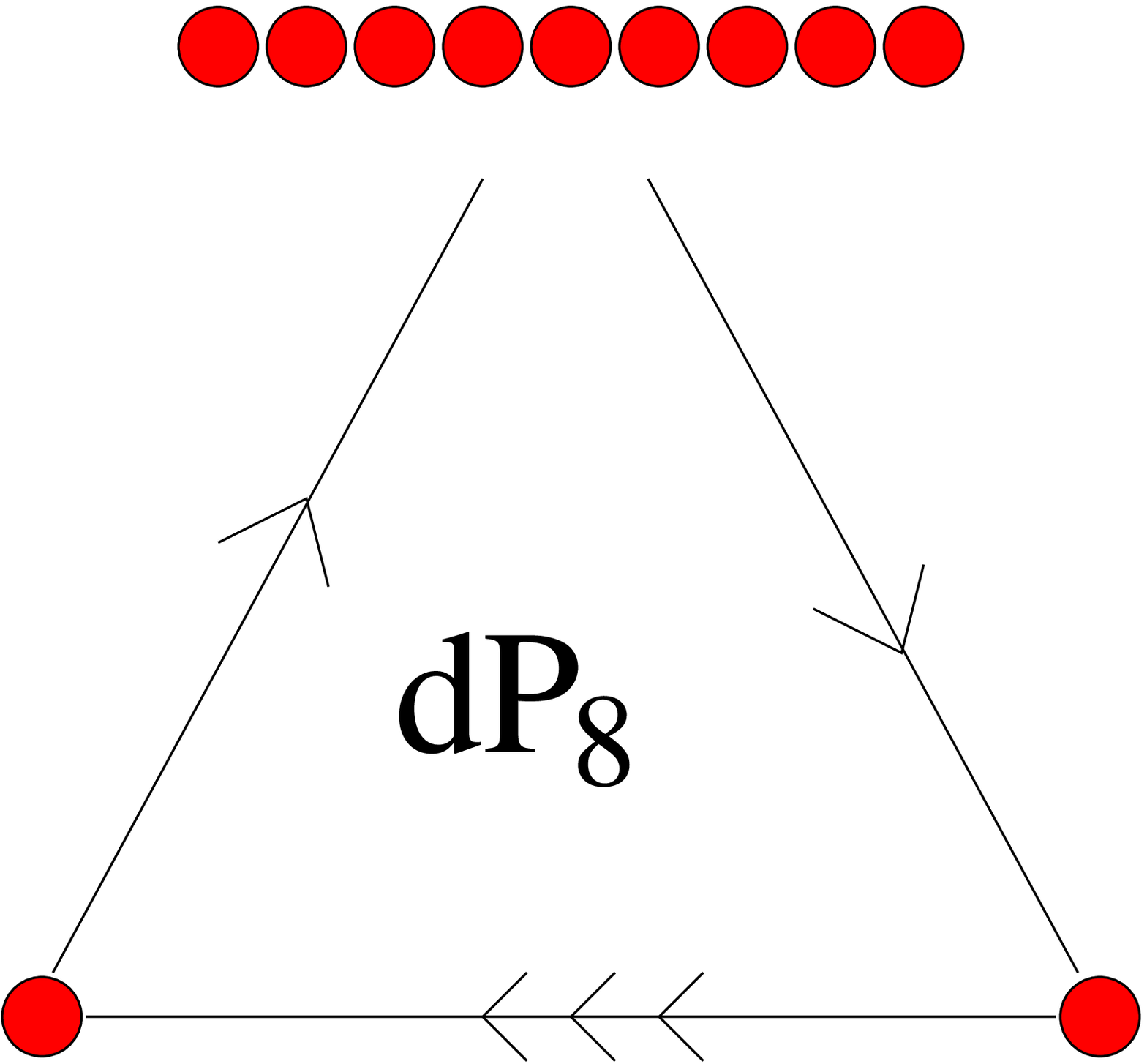}}&1&1&9  & 1&1&3& 3&3&1   &1 &$dP_8\;I$ \\
\hline
& 8 & 2 & 1    &  2 & 1 & 1 &    1 & 2 & 4    & 1  &  $dP_8\;II$ \\
\hline
& 2 & 3 & 6    &  1 & 1 & 1 &    3 & 2 & 1    & 1  &  $dP_8\;III$ \\
\hline
& 5 & 5 & 1    &  1 & 2 & 1 &    1 & 2 & 5    & 1  &  $dP_8\;IV$ \\
\hline\hline
\tsg{0.5cm}{1.5cm}\rb{-0.5cm}{\ig[height=1.4 cm]{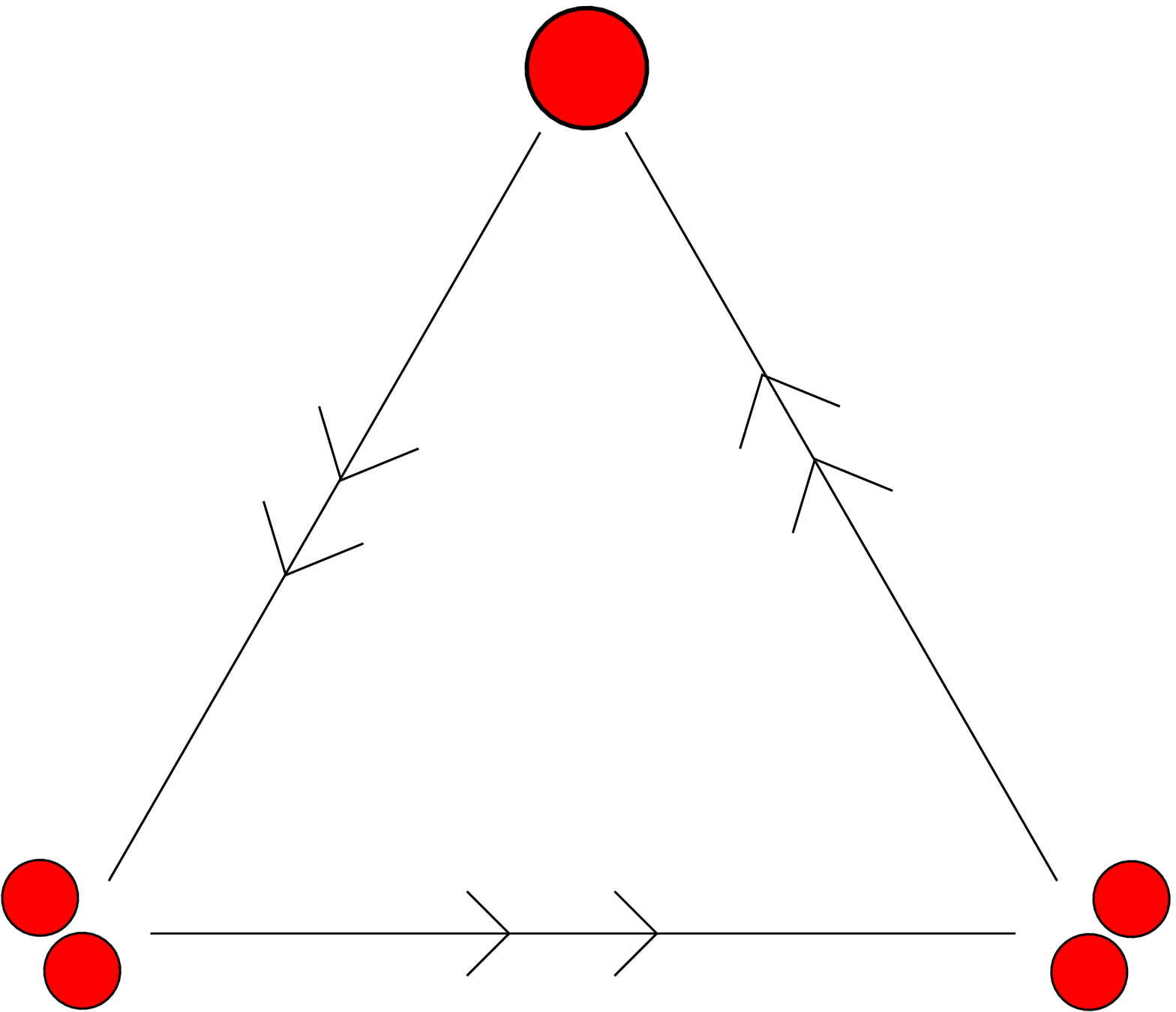}}&2&2&1& 2&2&2 & 1&1&2    & 4  & $sh\,dP_5$ \\
\hline
\tsg{0.5cm}{1.5cm}\rb{-0.5cm}{\ig[height=1.4 cm]{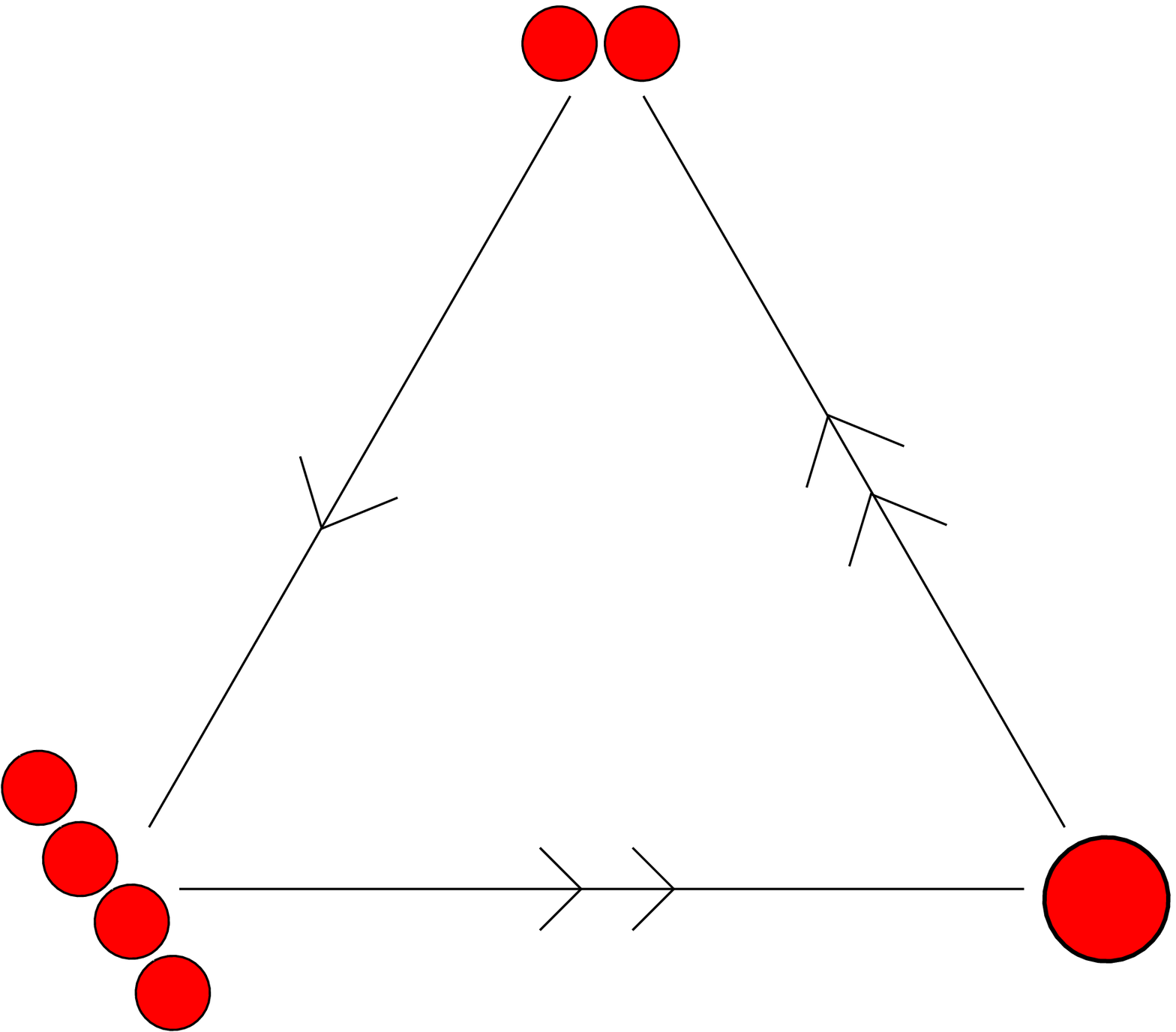}}&2&1&4& 2&1&2 & 2&2&1    & 2  & $sh\,dP_7$ \\
\hline
\end{tabular}
\caption{The complete list of the minimal models for $3$-block chiral quivers.}\label{list3b}
\end{center}
\end{table}
\end{center}
\newpage


\section{From one quiver to another: ``shrinking'' and ``orbifolding''}\label{quiveroperation}
In this section we discuss two general procedures that lead from a superconformal quiver to another superconformal quiver.
First we describe a well known procedure, orbifold, which has a very well known geometric meaning. Then we discuss another procedure, which we call ``shrinking''. The discussions in the two cases are quite similar; this is the reason why we consider also orbifolds and their effect on quivers. 


\subsection{Orbifolds}
Looking at the quiver diagrams of $\IF_0$ and $dP_5$ (Table \ref{list3b}), it is evident that the $dP_5$ quiver is the $\IF_0$ quiver with the order of the blocks $(\ag, \bg, \cg)$ multiplied by $2$ and the quiver numbers $(a, b, c)$ divided by $2$. This is a consequence of the fact that complex cone over $dP_5$ is a $\IZ_2$ orbifold of the cone over $\IF_0$:
\beq\label{orbF0}
\begin{CD}
\rb{-0.8cm}{ \ig[height= 2 cm]{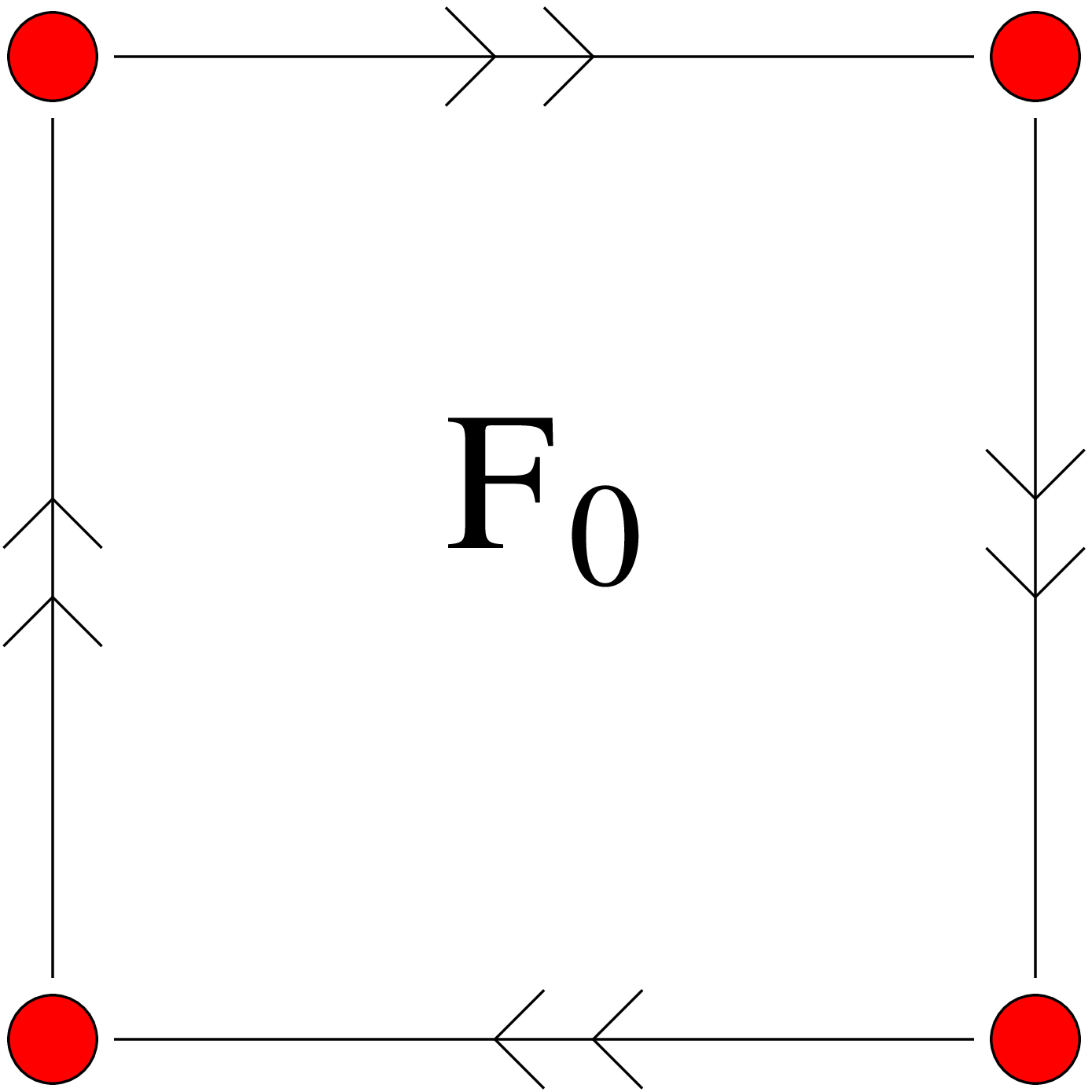}}
    \;\; @>orbifold \;\; \IZ_2>> \;\;
\rb{-0.8cm}{  \ig[height= 2 cm]{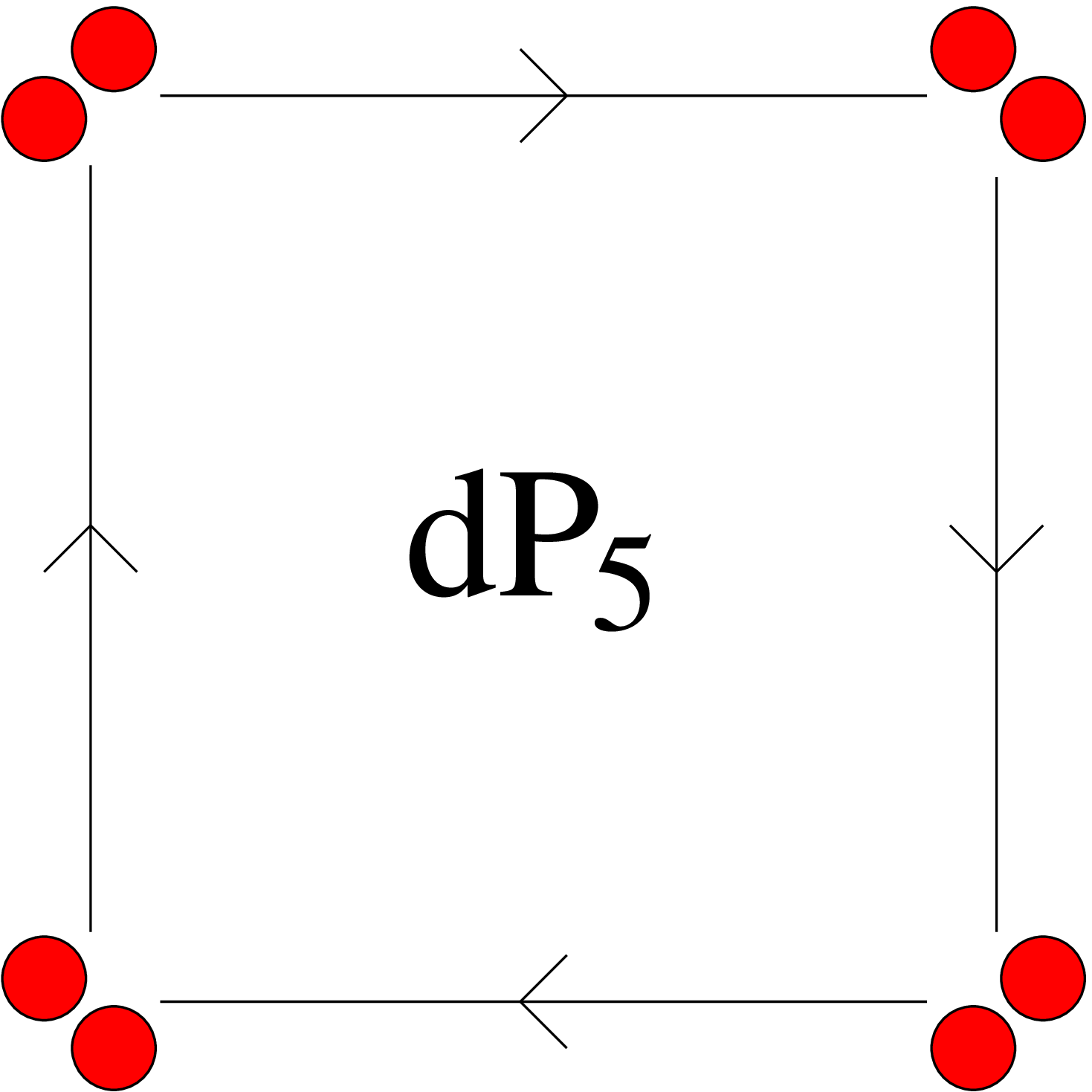}}
\end{CD}
\eeq
More precisely there is a point in the complex structure moduli space of $dP_5$ with this property \cite{Wijnholt:2002qz}, but the quiver diagram and the $r$-charges are not sensitive to a modification of the complex structure of the surface. What is sensitive is the precise form of the superpotential \cite{Wijnholt:2002qz}.

Another example of orbifolds inside the set of del Pezzo quivers is $dP_6$: a $\IZ_3$ orbifold of $dP_0$. In this case $(\ag, \bg, \cg)$ are multiplied by 3 and $(a, b, c)$ divided by $3$.
$$\begin{CD}
\rb{-0.8cm}{  \ig[height= 2cm]{delpezzo0.eps}}
 \;\; @>orbifold \;\; \IZ_3>> \;\;
\rb{-0.8cm}{  \ig[height= 2cm]{delpezzo6s.eps}}
\end{CD}$$
The last possible example coming from the list of three-block quiver of Table \ref{list3b} involves one of the three minimal models for del Pezzo $7$, $dP_7\,I$, and one of the ``new'' models, called in the Table \ref{list3b} $sh dP_5$. Also without knowing anything about the geometry dual to the $sh dP_5$ quiver, we can say that $dP_7\,I$ is a $\IZ_2$ orbifold of the $sh dP_5$ quiver:
$$\begin{CD}
\rb{-0.8cm}{  \ig[height= 2 cm]{shdp5.eps}}
 \;\; @>orbifold \;\; \IZ_2>> \;\;
\rb{-0.8cm}{  \ig[height= 2 cm]{delpezzo7s.eps}}
\end{CD}$$

The general procedure is clear: every time the elements of the quiver matrix $\mQ$ have a non trivial common divisor $d$, it is possible to obtain another quiver, corresponding to a $\IZ_d$ orbifold for the transeverse space. This daughter quiver has quiver matrix similar to the matrix of the parent theory, with dimension  multiplied by $d$ and elements divided by $d$. Each node is ``splitted'' in a $d$-node block. The (relative) ranks of the gauge groups and the $r$-charges do not change.

It is easy to verify that the daughter quiver satisfies the superconformal restrictions, if the parent quiver does. For each node in the original model there are $\nf_i$ incoming bifundamentals and $\nf_i$ outgoing bifundamentals. In the orbifolded model there are again $d \cdot (\nf_i/d) = \nf_i$ incoming bifundamentals and the same number of outgoing. The first consequence is that the ABJ anomalies cancel also in the orbifolded model. Second, since the relative numbers of flavors $\nf_i$ do not change, the beta function equations in the orbifolded quiver are satisfied by the same $r$-charges of the original quiver. The superpotential is a projection of the superpotential for the parent theory, so, since the $r$-charges do not change, the beta functions for the superpotential couplings still vanish.

It is important to note that if there is non trivial common divisor $d$ for an element of the Duality Tree, there will be the same common divisor for all the Duality Tree, so it is not important in which point of the Duality Tree the orbifold operation is performed. For instance for the $\IF_0$ and $dP_5$ quivers in (\ref{orbF0}) we chose two models slightly different from the ones shown in Table \ref{list3b}.

A common divisor $d$ for the quiver matrix indicates a non Abelian global symmetry $SU(d)$ for the model. For instance for $dP_0$ $d = 3$ and the superpotential of the $dP_0$ quiver can be chosen in such a way to preserve an $SU(3)$ global symmetry; for $\IF_0$ $d = 2$ and the global symmetry is $SU(2) \times SU(2)$, containing $SU(2)$. Finding a superconformal quiver with common divisor greater than $3$ would lead to difficulties for the existence of a string description of the quiver. The reason is that the global non Abelian symmetries of the conformal quivers correspond holographically to isometries of the transverse space $X_5$\footnote{There can be baryonic symmetries not coming from isometries of $X_5$ \cite{Berenstein:2002}, but these are always Abelian. The recently discovered non Abelian extension \cite{Franco:2004} holds in the infinite coupling limit.}, where the holographic dual is given by Type IIB strings on $AdS_5 \times X_5$. This isometry group cannot be larger then $SU(4) = SO(6)$, the maximal symmetry group for a compact 5-dimensional real manifold. With a common divisor $4$ we would have an isometry group containing the $SU(4)$ factor and also the $U(1)$ $R$-symmetry, which is too big.

The process just discussed does not describe the more general orbifolds, since it leaves the conditions of superconformality and chirality untouched. More general geometric orbifolds are in correspondence with discrete subgroups of the isometry group of the transverse space. In the two del Pezzo examples given above the discrete subgroup is in the center of the isometry group. This suggests that also for the $sh dP_5$ quiver the superpotential preserves at least an $SU(2)$ global symmetry. Since the superpotential for the orbifolded theory can be easily obtained from the superpotential for the parent theory, one could think that it is possible to reconstruct the moduli space (and thus the geometry) from the geometry of $dP_7$. Unfortunately the superpotential for the quiver $dP_7$ is still not known.

\subsection{Shrinking}
In the last paragraph we saw that if \emph{all} the numbers of the intersection matrix share a non trivial common divisor it is possible to obtain another superconformal quiver by the procedure of orbifolding.

In fact there is another similar procedure. Suppose we have a node (say node $1$) such that there is a non trivial common divisor $d$ for the number of arrows connected to that node: $\mQ_{1 i} = d \, q_{1 i}$\footnote{We observe that this condition is invariant under Seiberg Duality, if it holds in a point of the Duality Tree it holds in all the Duality Tree.}. Also in this case it is possible to ``split'' the quiver; the steps are the following:
\begin{itemize}
\item split node $1$ in a block of $d^2$ nodes;
\item divide by $d$ the multiplicity of the arrows connected to node $1$;
\item divide by $d$ the rank of the group $U(N_1)$ associated to node $1$;
\item leave untouched the rest of the quiver diagram.
\end{itemize}
We call the inverse of this splitting procedure ``shrinking of a block of $d^2$ nodes'':

\beq\label{shfig}
\begin{CD}
\rb{-1.8cm}{  \ig[height=4 cm]{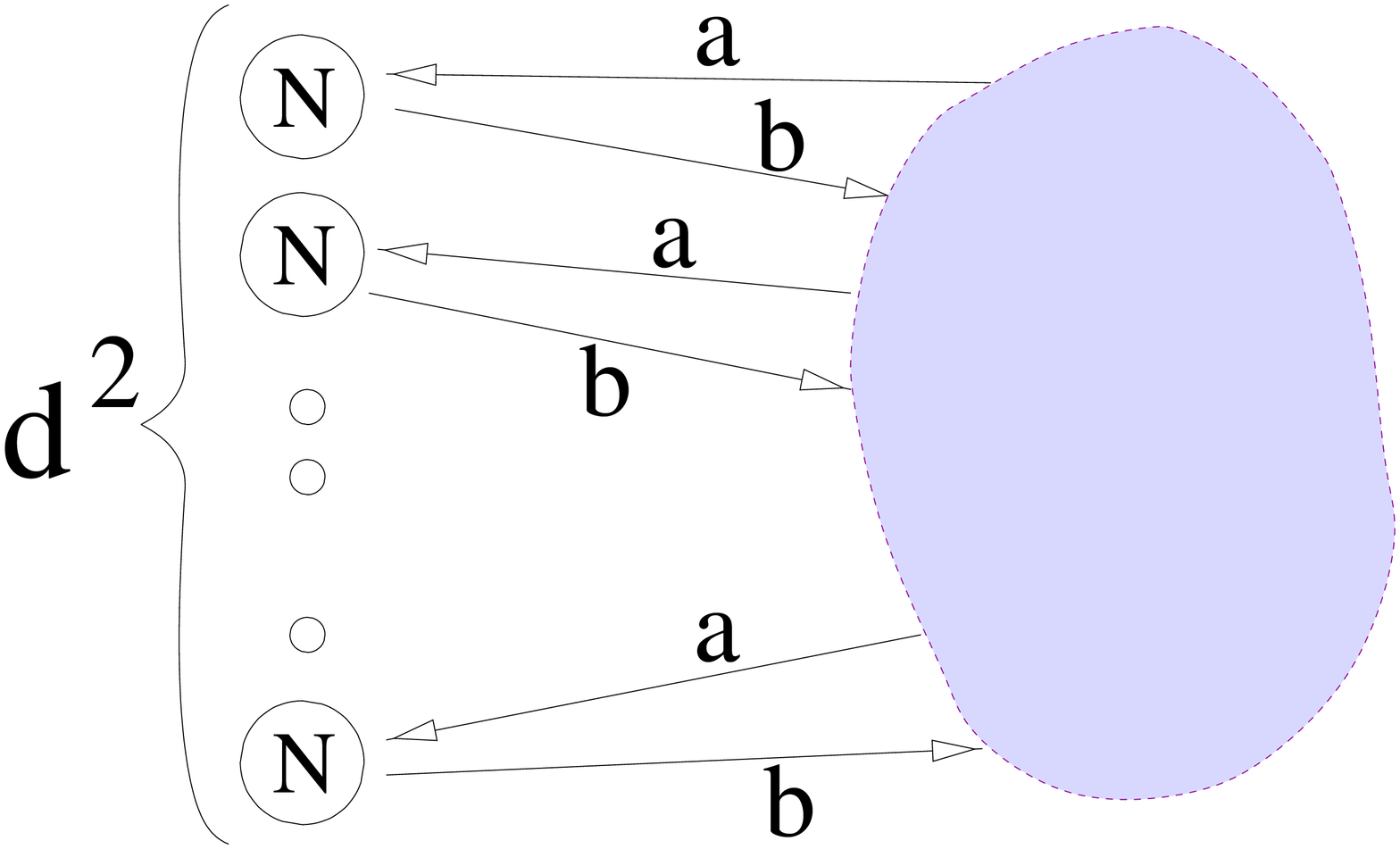}}
 \;\; @>shrinking>> \;\;
\rb{-1.8cm}{  \ig[height=4 cm]{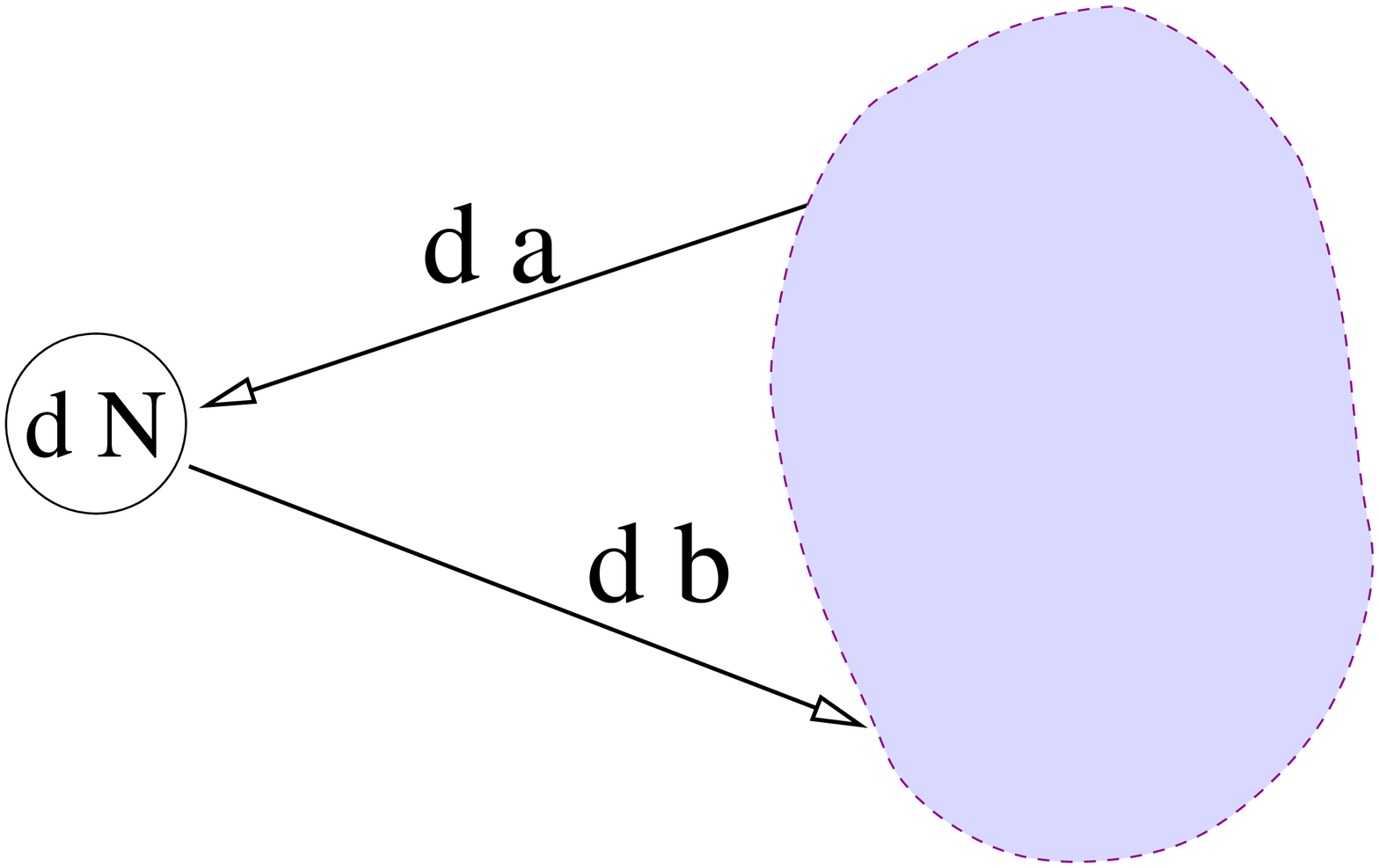}}
\end{CD}
\eeq

As for the orbifold it possible to give a simple proof of the fact that the shrunk model satisfies the superconformal conditions. We discuss this at the end of this paragraph, after having given some examples, taken by the $3$-block list of Table \ref{list3b}.

Every quiver containing a block with $4$ or more nodes can be shrunk, and the result is a new different quiver. This is true also for non-chiral superconformal quivers, even if we only discuss chiral examples.

One example is the shrinking of $9$ nodes of the $3$-block $dP_8$ quiver with $(\ag, \bg, \cg) = (1 ,1, 9)$; this leads to the $dP_0$ quiver, with ($\ag, \bg, \cg) = (1 ,1, 1)$:
\beq\label{sh3dp8}
\begin{CD}
\rb{-0.8cm}{  \ig[height=2 cm]{delpezzo8s.eps}}
 \;\; @>shrinking>> \;\;
\rb{-0.8cm}{  \ig[height=2 cm]{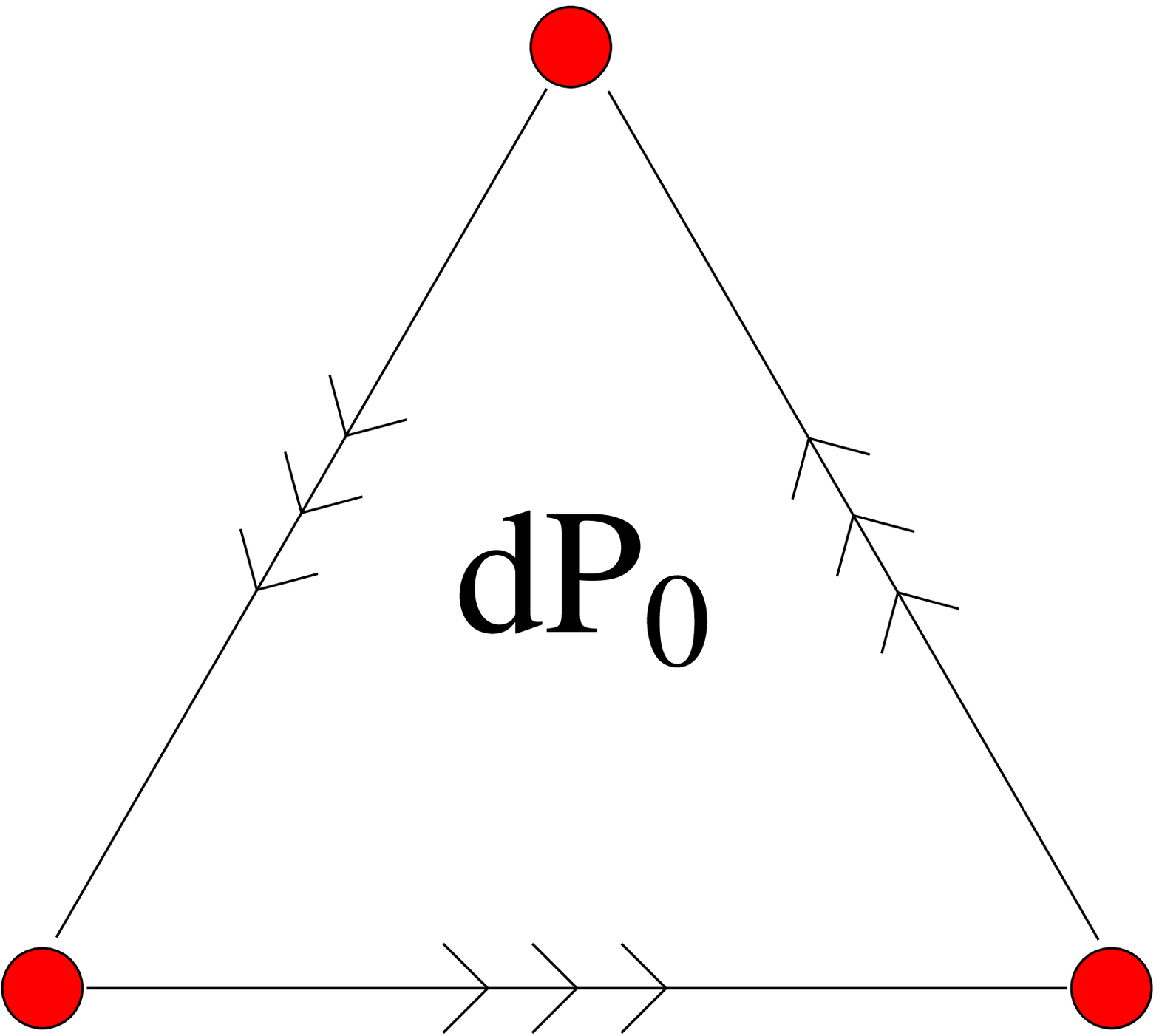}}
\end{CD}
\eeq
Note that this model of $dP_8$ has zero anomalous dimensions and therefore is a candidate to be an orbifold model. Indeed this quiver is precisely that of the orbifold $\IC^3/\Delta(27)$ \cite{Hanany:1998sd,Feng:2000mw}.
It is interesting that the two non del Pezzo $3$-block quivers can be obtained from del Pezzo quivers by shrinking a block of four nodes: it is possible to shrink $4$ nodes of $dP_5$ and of $dP_7$.
$$\begin{CD}
\rb{-0.8cm}{  \ig[height=2 cm]{delpezzo53bs.eps}}
  @>shrinking>>
\rb{-0.8cm}{  \ig[height=2 cm]{shdp5.eps}}
\end{CD}$$
$$\begin{CD}
\rb{-0.8cm}{  \ig[height=2 cm]{delpezzo7s.eps}}
  @>shrinking>>
\rb{-0.8cm}{  \ig[height=2 cm]{shdp7.eps}}
\end{CD}$$
The model $sh dP_7$ still contains a block with four nodes, so it can be shrunk again. The result is the $\mathbb{F}_0$ quiver. It is thus possible to construct the following chain of shrinking and orbifolds:
$$\begin{CD}
F_0  @>orbifold>>    dP_5  @>shrinking>> sh dP_5  @>orbifold>>
dP_7    @>shrinking>>    sh dP_7     @>shrinking>>   F_0
\end{CD}$$
This succession of quivers in our opinion suggests that a string interpretation of the quivers $sh dP_5$ and $sh dP_7$ should be possible.

We also note that shrinking the block with $8$ nodes in the $dP_8$ quiver with $(\ag, \bg, \cg) = (1 ,2, 8)$ leads again to the $sh dP_5$ quiver:
\beq\label{shdp8}
\begin{CD}
dP_8   @>shrinking>>sh dP_8 @>shrinking>>   sh dP_5
\end{CD}
\eeq

It is possible to shrink a quiver with a block with more than $4$ nodes: first one divides the block in a block of $4$ nodes and in another block with the remaning nodes, then one shrinks the $4$-node block. Applying this procedure to a $3$-block quiver one obtains a $4$-block quiver. For example the intermediate quiver $sh dP_8$ in (\ref{shdp8}) is a $4$-block quiver; we discuss some other examples in the last paragraph of this section.

\subsection*{The shrunk models are conformal}
Here we discuss the reason why the shrinking procedure is non trivial: if the original quiver is superconformal also the shrunk one satisfies the superconformal constraints. We recall that constructing supersymmetric quivers is very easy (it is enough to cancel the gauge anomalies), while imposing that an interacting RG fixed point exists (superconformal quiver) is much more difficult: it is necessary to exactly cancel the beta functions for all the couplings of the theory.

The proof is easy and it is very similar to the arguments given in the last subsection for orbifolds. If we show that the relative number of flavours $\nf$ does not change for each node (for the new node this means that it has the same $\nf$ of the nodes in the shrunk block) the proof is almost finished: the ABJ anomalies cancel and the gauge couplings beta functions vanish with the same $r$-charges of the original theory; with the same $r$-charges also the superpotential couplings beta functions vanish, since the possible loops in the shrunk theory are in correspondence with loops of the original theory. So we have to study the relative number of flavours $\nf$.

Let's analize first node $1$ (the shrunk one). The nodes in the old $d^2$-nodes block have (as in equation (\ref{nfdef})) 
\beq\label{nfblock} 
\nf_{block} = \frac{\sum_{i > d^2} |\mQ^{old}_{1 i}| N^{old}_i}{2 \, N^{old}_1}\;,
\eeq
where we label the nodes outside the $d^2$-nodes block with integers starting from $d^2 + 1$ (both in the old and in the new quiver). The new shrunk node has
\beq\label{nfshrunk}
\nf_{shrunk} = \frac{\sum_{i > d^2} |\mQ^{new}_{1 i}| N^{new}_i}{2 \, N^{new}_1}\;.
\eeq
Taking into account that
\beq\label{shprop} 
\left\{ \begin{array}{l}
\mQ^{new}_{1 i} = d \, \mQ^{old}_{1 i}\\
N^{new}_1 = d \, N^{old}_i \phantom{aaaaa}  if \;\; 1 \leq i \leq d^2\\
N^{new}_i = \; N^{old}_i \phantom{iiaaiaa}   if \;\;\; i > d^2\;,
\end{array}\right.
\eeq
it is clear that (\ref{nfblock}) and (\ref{nfshrunk}) are equal.

Now we consider $\nf$ for a generic node labelled by $i > d^2$ (in the clouds of (\ref{shfig})).
\beq\label{nfold}
\nf_{old}   = \frac{\sum_{j \neq i} |\mQ^{old}_{i j}| N^{old}_j}{2 \, N^{old}_i} = 
              \frac{ \sum_{j = 1}^{d^2} |\mQ^{old}_{i j}| N^{old}_j}{2 \, N^{old}_i} +
              \frac{ \sum_{d^2 < j \neq i} |\mQ^{old}_{i j}| N^{old}_j }{2 \, N^{old}_i}\;,
\eeq
The first addendum on the $r.h.s.$ represents the $d^2$ equal contributions from the $d^2$-nodes block and, keeping track of (\ref{shprop}), can be rewritten as
\beq\label{nfa}
\frac{ \sum_{j = 1}^{d^2} |\mQ^{old}_{i j}| N^{old}_j}{2 \, N^{old}_i} = 
\frac{d^2 |\mQ^{old}_{i 1}| N^{old}_1}{2 \, N^{old}_i} = \frac{|\mQ^{new}_{i 1}| N^{new}_1}{2 \, N^{new}_i}\;.
\eeq
The second addendum in (\ref{nfold}) represents contributions coming from nodes inside the cloud, it obviously satisfies
\beq\label{nfb}
\frac{ \sum_{j > d^2} |\mQ^{old}_{i j}| N^{old}_j }{2 \, N^{old}_i} = 
\frac{ \sum_{j > d^2} |\mQ^{new}_{i j}| N^{new}_j }{2 \, N^{new}_i} \;.
\eeq
The relative number of flavours $\nf$ for a generic node in the shrunk quiver is given by
\beq\label{nfnew}
\nf_{old}   = \frac{\sum_{j \neq i} |\mQ^{new}_{i j}| N^{new}_j}{2 \, N^{new}_i} = 
              \frac{  |\mQ^{new}_{i 1}| N^{new}_1}{2 \, N^{new}_i} +
              \frac{ \sum_{j > d^2} |\mQ^{new}_{i j}| N^{new}_j }{2 \, N^{new}_i}\;.
\eeq
It is clear that (\ref{nfnew}) is equal to the sum of (\ref{nfa}) and (\ref{nfb}), so it is equal to (\ref{nfold}).

In conclusion we have shown that the relative number of flavours $\nf$ are the same in the original and in the shrunk model, completing the proof that the shrunk quiver is a consistent superconformal field theory. Once again we would like to remark that this fact does not depend on the chirality of the quiver, a property that indeed wasn't used.

\subsection{Some general properties of shrunk quivers}
In this subsection we would like to make some general comments on the properties of the shrunk models. The most interesting thing to understand is if these new superconformal field theories arise as the low energy excitations of a stack of D3 branes in Type IIB string theory, as the original models do. As discussed in Section \ref{Wandgeom} if one finds a superpotential leading to a $3$-complex dimensional cone the answer should be positive. This is however not necessary, it could happen that the string dual has non trivial R-R fluxes, leading to a potential for the D3 branes and effectively decreasing the dimension of the moduli space of the gauge theory. It is also possible that the string dual has non trivial NS-NS fluxes. In the case of orbifolds this is called Discrete Torsion. NS-NS fluxes in the cases analised in the literature always lead to a decrease in the number of nodes in the quiver: this is similar to our shrinking procedure and suggests that shrinking generically involves (on the geometric side) turning on NS-NS forms. It is possible that discrete R-R fluxes, which can have the generic name discrete torsion as well, will have a similar effect.

Studying all the possible moduli spaces as a function of the superpotential is not an easy task, however this search can be simplified by restricting to superpotentials that preserve a global symmetry $SU(d)$. This symmetry for the shrunk models arise in way similar to the orbifold: the multiple legs connected to the shrunk node can be charged under a non Abelian symmetry $SU(d)$, since there is precisely a common divisor $d$. For instance in (\ref{sh3dp8}) $3$ incoming and $3$ outgoing legs are connected to the shrunk node and in the shrunk quiver, which is del Pezzo $0$, there exist a (unique) superpotential preserving an $SU(3)$ global symmetry. This symmetry, together with the $U(1)$ $R$-symmetry, corresponds to the isometries of the transverse space, which, in this case, is simply the orbifold $\IC^3/\IZ^3$.\\
Also for the shrinking we note that finding a non abelian global symmetry bigger than $SU(3)$ would prevent the existence of a holographic dual; interestingly in our classifications we never find a superconformal quiver with a (local) common divisor greater than $3$, or, equivalently, a superconformal quiver with a block with $16 = 4^2$ or more nodes.

Now we would like to discuss some aspects anticipated in Section \ref{unitarity}. First of all we note the rank of the quiver matrix does not change\footnote{In the case of non-chiral quiver this statement is still meaningful, if we define the rank of the quiver matrix to be the rank of the antisymmetric part of the matrix describing the number of arrows in the quiver. With this definition the rank of the conifold and the generalized conifolds vanishes, since these model are completely non-chiral, and the antisymmetric part is a zero matrix.}; the reason is quite simple. Consider the upper triangular part of the antisymmetric quiver matrix. The procedure inverse to shrinking substitutes a column (say the last) with $d^2$ identical columns:
\beq\begin{CD}
\left(\begin{array}{ccccc}
  0 &\; * \;&\ldots&\;\; * \;&d\,\mQ_{1}\\
    &\; 0 \;&      &\;\vdots &\vdots \\
    &       &\ddots&\;\; * \;&\vdots \\
    &       &      &\;\; 0 \;&d\,\mQ_{n}\\
    &       &      &         &   0     
\end{array}\right) 
 \;\; @<shrinking<< \;\; 
\left(\begin{array}{ccccc}
  0 &\; * \;&\ldots&\;\; *  \!&\overbrace{\mQ_{1}\ldots\mQ_{1}}^{d^2}\\
    &\; 0 \;&      &\;\vdots\!&\vdots \phantom{aaaaa} \vdots     \\
    &       &\ddots&\;\; *  \!&\vdots \phantom{aaaaa} \vdots    \\
    &       &      &\;\; 0  \!&\mQ_{n}\ldots\mQ_{n}\\
    &       &      &        \!& 
 
\phantom{iaa} \left.\begin{array}{ccc}   
  0 &\, \ldots \,& 0 \\
    &   \ddots   & \vdots\\
    &            & 0
\end{array}\right\}d^2

\end{array}\right)\;.
\end{CD}\eeq
It follows that, considering the full antisymmetric matrix, the rank does not change. In our cases thus, both the del Pezzo quivers and the shrunk del Pezzo quivers have rank $2$.

Another important feature of the superconformal field theory is the invariance under shrinking of the two gravitational central charges $c = a$\footnote{One could object that the central charges of del Pezzo 8 and del Pezzo 0 quivers are different (the first is $9$ times bigger than the latter), and this is in contrast with the fact that shrinking $9$ nodes of del Pezzo $8$ leads to del Pezzo $0$, as in (\ref{sh3dp8}). The explanation is that the ranks of the gauge groups in del Pezzo $8$ are $( 3 N, 3 N, N, \ldots, N )$, as shown in Table \ref{list3b}, so shrunk del Pezzo 8 has ranks $( 3 N, 3 N, 3 N )$. Formula (\ref{canda}) thus explains the factor of $9$. The same observation applies to the ``equality'' between $\IF_0$ and del Pezzo $7$ shrunk two times.}. Recalling formulae (\ref{AFGJ}) and the fact the for superconformal quivers $tr(R) = 0$ we have to show that $tr(R^3)$ is invariant. We can recognize in formula (\ref{traceR})
\beq
tr R^{3} =\hspace{-0.4 cm} \sum_{G\in gauge\,groups}\hspace{-0.4 cm} dim[G] \;  + \hspace{-0.4 cm}
\sum_{B\in bifundamentals} \hspace{-0.4 cm} dim [\mathcal{R}_B] (r_M - 1)^{3}
\eeq
$4$ different contributions: the gauge groups inside the clouds of (\ref{shfig}), the bifundamental fields inside the clouds, the gauge groups forming the $d^2$-node block (becoming the shrunk node) and the contribution coming from the bifundamental connecting the $d^2$-node block to the rest of the quiver. Each of these $4$ contributions is invariant.\\
For the first two summands this is clear, since the quiver does not change inside the clouds and all the $r$-charges are the same.\\
The third summand is the initial block with $d^2$ nodes of rank $N$. Its contribution to $tr(R^3)$ is $d^2\,N^2$. The contribution coming from the shrunk node of rank $d\,N$ is $(d\,N)^2$ too, so also this part is invariant.\\
The fourth summand before shrinking contributes as
\beq
d^2\,N \sum_{j > d^2} |\mQ^{old}_{1 j}| N^{old}_j\,(r_{1 j} - 1)^{3}\;;
\eeq
after shrinking it becomes
\beq
( d\,N ) \sum_{j > d^2} |\mQ^{new}_{1 j}| N^{new}_j\,(r_{1 j} - 1)^{3}\;.
\eeq
This two expressions are easily seen to be equal recalling that, (\ref{shprop}), $\mQ^{new}_{1 i} = d \, \mQ^{old}_{1 i}$ and $N^{new}_i = \; N^{old}_i$ if $i > d^2$.

In conclusion all of the $4$ contributions are the same in the original and in the shrunk models.

We are now able, as anticipated, to explain the relations (\ref{KSQ-nodes}), relating $K^2$ and the number of nodes. The starting point is (\ref{canda}):
\beq \label{candab}
c = a = \frac{27 N^2}{4\,K^2}\;.
\eeq
Since the gravitational central charges do not change under shrinking, also $K^2$ is invariant\footnote{This is true if after the shrinking there isn't a new common divisor for the rank of the gauge groups, so this observation does not applyies to \mbox{del Pezzo $8$ $\rightarrow$ del Pezzo $0$} or \mbox{del Pezzo $7$ $\rightarrow$ $\IF_0$}.}. The shrunk model have $3$ nodes less with respect to the original ones (a block of $4$ nodes becomes just $1$ node), so from the relation
\beq 
K^2  = 12 - (\ag + \bg +\cg) \phantom{aaaaaa}  \text{for del Pezzo quivers}
\eeq
it follows that
\beq 
K^2  = 9  - (\ag + \bg +\cg) \phantom{aaaaaa}    \text{for del Pezzo quivers with $4$ nodes shrunk.}
\eeq

\subsection{New superconformal chiral quivers}
Applying the procedure of shrinking it is possible to contruct new rank-$2$ superconformal quivers. We have to look for blocks with $4$ or more nodes. The smallest (in the sense of the total number of nodes) quiver could be a three-block quiver with $6$ nodes, $( \ag, \bg, \cg ) = (1, 1, 4)$, but such a quiver does not exist, as follows from the classification of $3$-block quivers given in the appendix.

With $7$ nodes it is possible to shrink $dP_4$ quiver, which has $( \ag, \bg, \cg ) = (1, 1, 5)$. In order to do this it necessary to ``open'' the block with $5$ nodes in a block of $4$ nodes and a block with $1$ node, then shrink the block with $4$ nodes. The result is a $4$-block chiral superconformal quiver:
$$\begin{CD}
\rb{-0.8cm}{  \ig[height=2 cm]{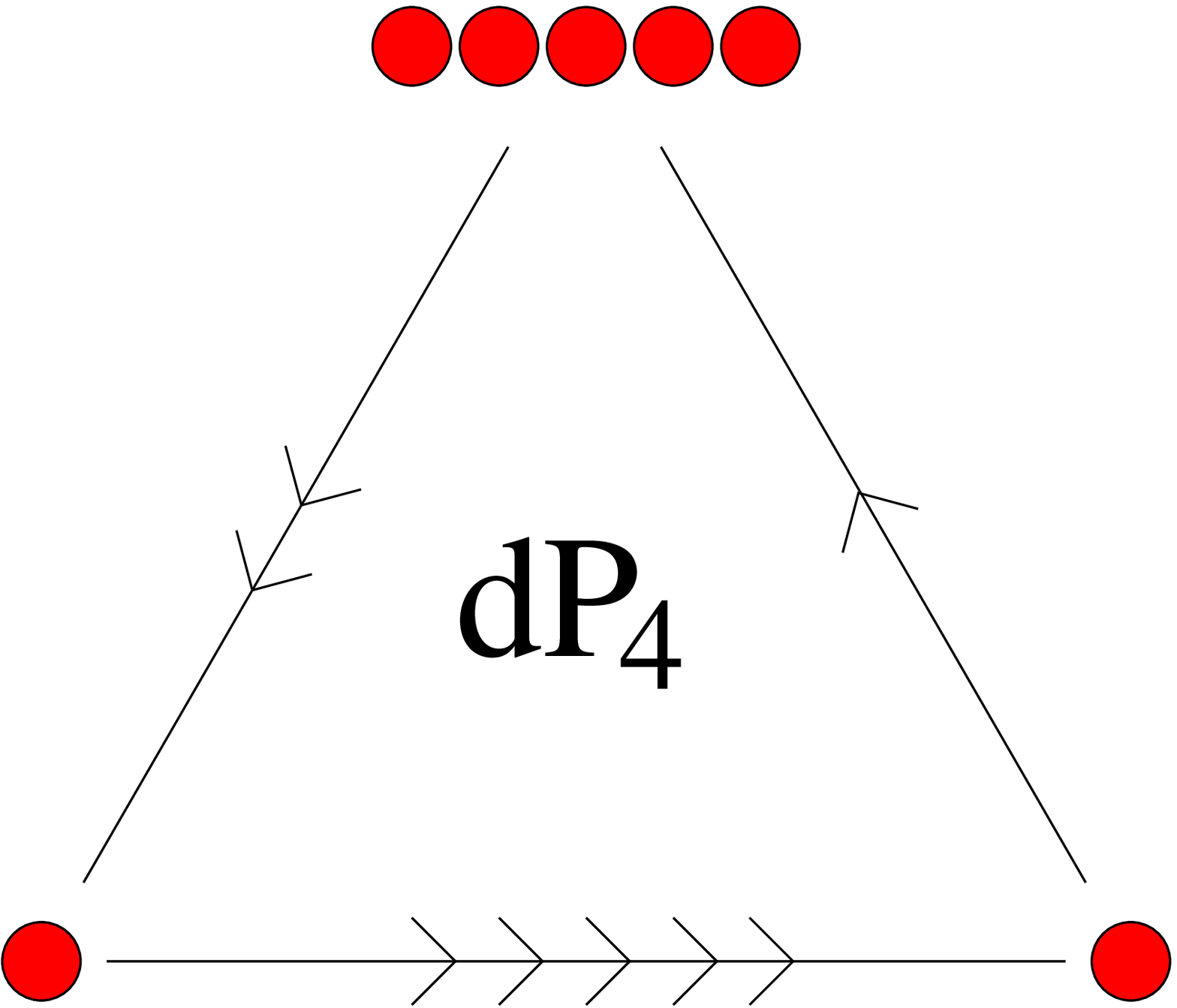}}
 \;\; \sim \;\;
\rb{-0.8cm}{  \ig[height=2 cm]{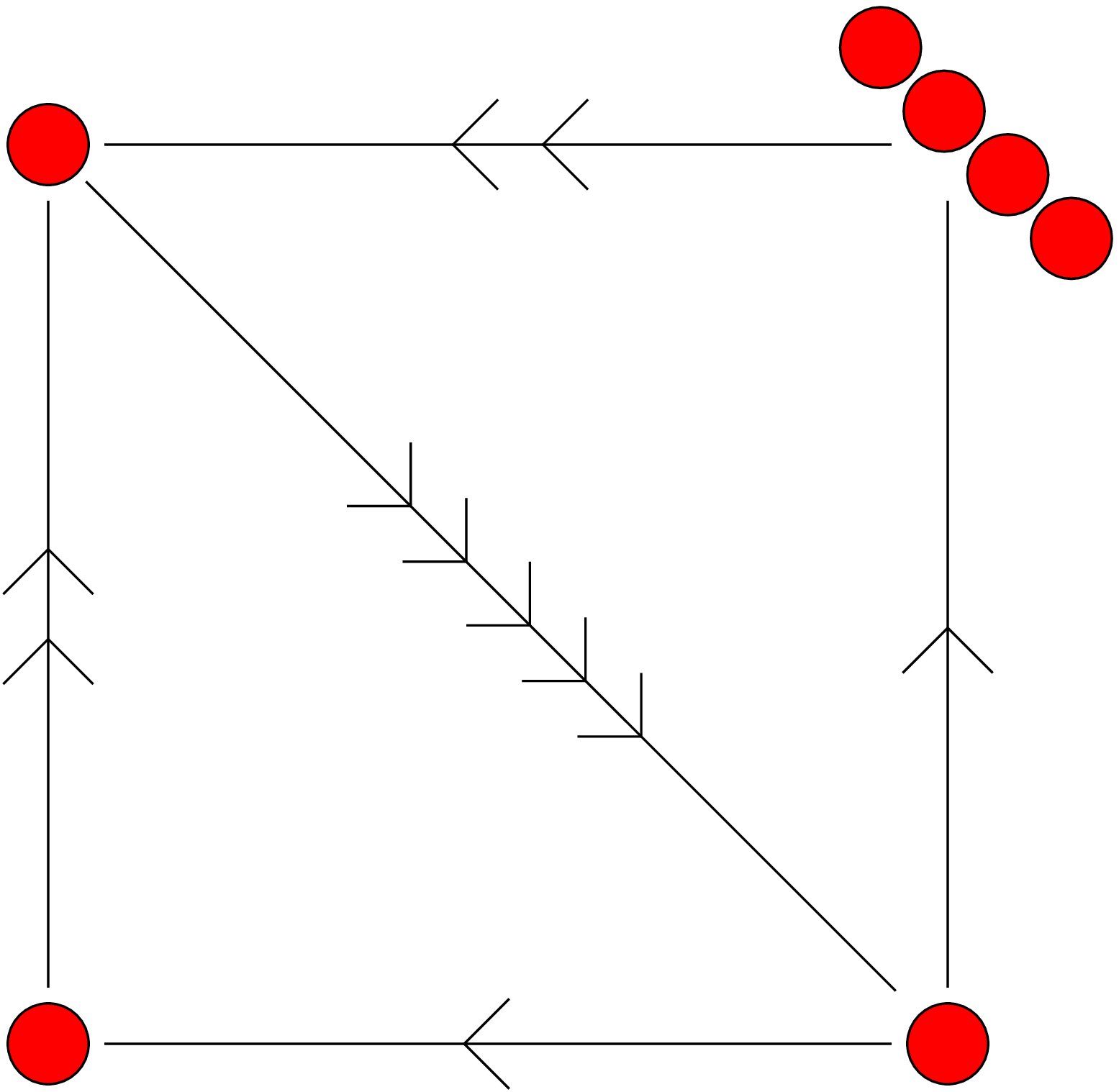}}
 \;\; @>shrinking>> \;\; 
\rb{-0.8cm}{  \ig[height=2 cm]{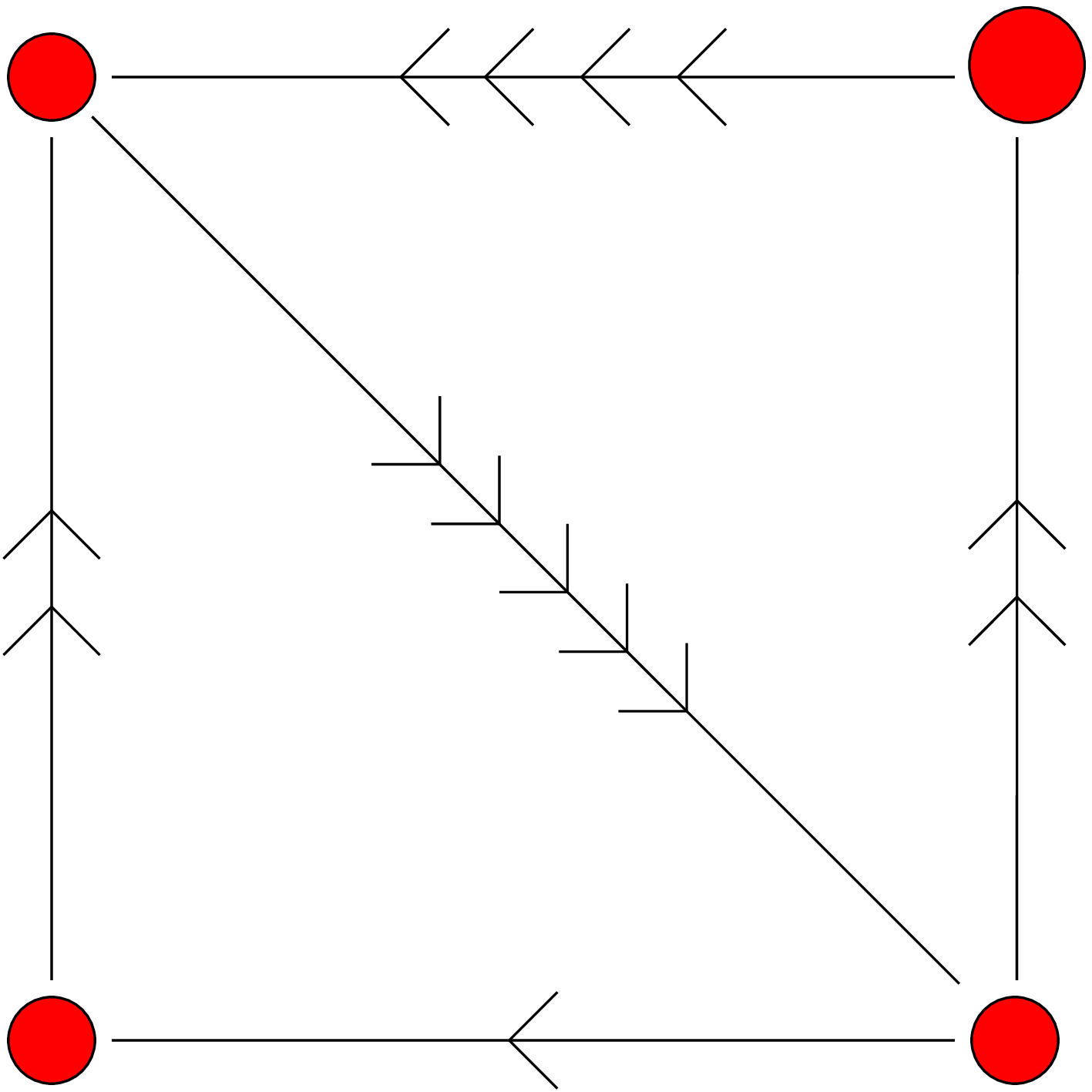}}
\end{CD}$$

With $8$ nodes there is just the $dP_5$ quiver to shrink, leading to the already discussed three-block $5$-node quiver $sh dP_5$.

With $9$ nodes we can shrink $dP_6$, model II, with $( \ag, \bg, \cg ) = (2, 1, 6)$. In order to do this it necessary to open the block with $6$ nodes then shrink the resulting block with $4$ nodes:
$$\begin{CD}
\rb{-0.8cm}{  \ig[height=2 cm]{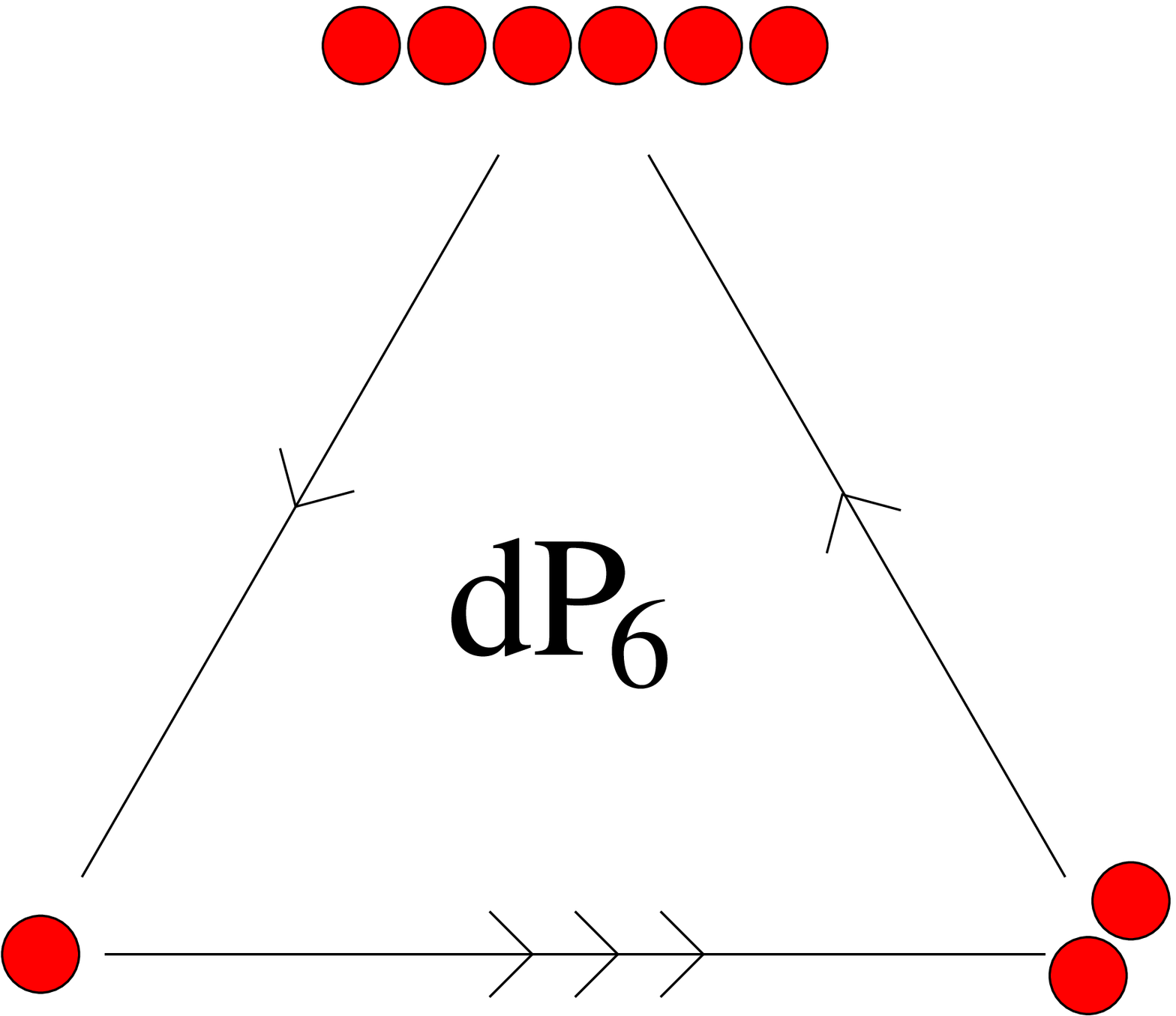}}
  \;\;  @>shrinking>> \;\; 
\rb{-0.8cm}{  \ig[height=2 cm]{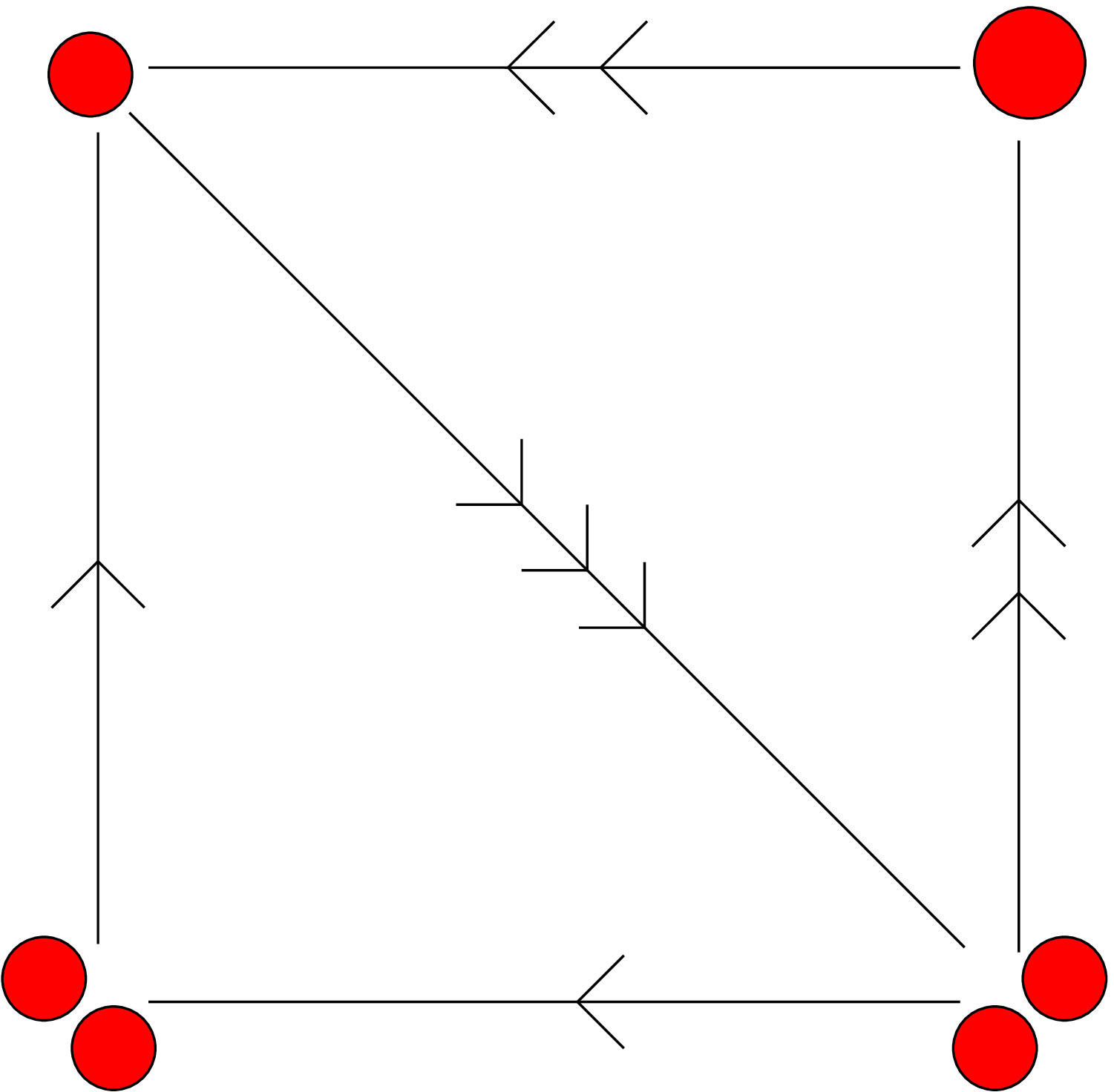}}
\end{CD}$$

This procedure can be applied also to the quivers $dP_7$ and $dP_8$. In these two cases there are however some differences.\\
First, it is possible to obtain inequivalent quivers from equivalent original quivers (equivalent means related by successive Seiberg Dualities). The reason is that after the shrinking there are less nodes, so there is less freedom for the Seiberg Dualities, and the result is that it is impossible to Seiberg Dualize the shrunk models between each other.\\
Second, it is possible to shrink the $dP_7$ and $dP_8$ quivers two times. In the case of $dP_7$ this leads to a quiver identical to the $F_0$ quiver. In the case of $dP_8$ the double shrinking leads to shrunk-del Pezzo $5$ or to new $5$-nodes quivers.

We will not describe in detail all the possible ``shrunk'' chiral superconformal quivers, since the procedure for contructing them is by now clear. In the next section we will analize the case of $4$-block chiral quivers. The superconformal restrictions lead also in this case to a set of diophantine equations that classify all the possible models. A computer-based search of solutions to these diophantine equations leads to the result that all non del Pezzo models can be obtained applying the shrinking procedure to a del Pezzo quiver.


\section{Four-block chiral quivers}\label{fourblock}
In this section we analyze in detail the supersymmetric chiral quivers that can be organized in a four-block structure. We study  the constraints on the matter content of the quiver arising by anomaly cancellation and by conformal invariance, obtaining (like in the 3-block models) a diophantine equation which classifies all possible models.

In this class of superconformal quivers there are of course models that can be Seiberg Dualised to $3$ blocks, which have been analised in Section \ref{threeblock}. In the previous section we saw that it is possible to contruct, beyond the quivers that admit a $3$-block structure, ``new'' four block models arising from the operation of shrinking $4$ nodes of the del Pezzo $n$ quivers, with $4 \leq n \leq 8$. Again we notice that it is not possible to construct the quivers of del Pezzo $1$ and del Pezzo $2$.

The main result of this section is that \emph{all} the chiral $4$-block quivers are of this type: there are only the del Pezzo and the shrunk-del Pezzo quivers. This result is obtained studying the possible solutions of the $4$-block diophantine equation. In this case, however, we have not been able to give an analytic proof of this fact. Our assertion is based on a search of solutions done on the computer. Looking for ``roots'' of Duality Trees one finds (similarly to the $3$-block case) solutions given by quiver matrices formed by ``small'' integers, then, increasing the size of these integers, one does not find solutions anymore. In the $3$-block case we were able to give a detailed analitic proof of the absence of other solutions (beyond del Pezzo and shirinked-del Pezzo). In the $4$-block case we have to rely on computer simulations, but we think that this number theoretic result is quite solid.

This result implies that all the physical properties discussed in Section \ref{threeblock} for $3$-block quivers (unitarity and asymptotic freedom) hold also for all the $4$-block models. The reason is that they are valid for del Pezzo quivers and the procedure of shrinking leave these properties untouched.

Recalling that the quiver matrix of all the $3$- and $4$-block models has rank $2$, our classification results lead naturally to the conjecture that \emph{all the rank-$2$ chiral quivers are del Pezzo quivers or can be obtained by shrinking one or two blocks in a del Pezzo quiver}. This conjecture implies in particular that rank-$2$ chiral quivers have at most $11$ nodes, since the del Pezzo quiver with the higher number of nodes, del Pezzo $8$, has precisely $11$ nodes.

\subsection{Structure of the quiver diagram}
First of all we have to understand the topological structure of the quiver diagram, i.e. the possible direction of the arrows. We will see that for the case of four blocks there is just one possibility.

\begin{figure}[!h] 
\begin{center}
\ig[height=6.8 cm]{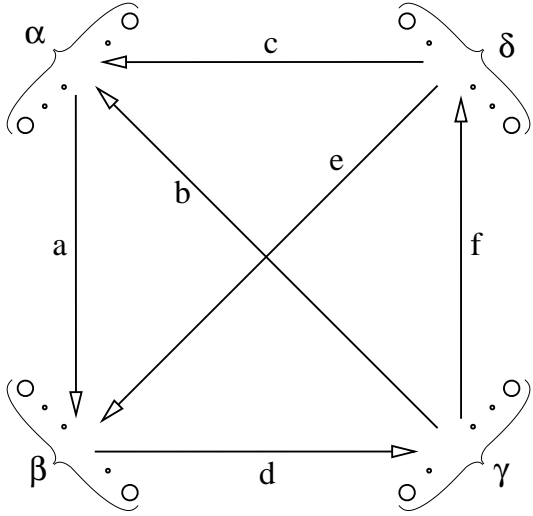}
\caption{4-block  chiral quiver diagram.}\label{figquiver_4block}
\end{center}
\end{figure}

 For each node in the quiver diagram cancellation of the gauge anomalies tells that it is forbidden to have all the $3$ arrows incoming or all the $3$ arrows outgoing. There are thus two possible types of nodes:
\begin{itemize}
\item first kind: $2$ incoming arrows and $1$ outgoing arrow,
\item second kind: $1$ incoming arrow and $2$ outgoing arrows.
\end{itemize}
Since each arrows is outgoing from a block and ingoing into another there has to be exactly $2$ blocks of the first kind and $2$ blocks of the second kind.\\
Now, if the two outgoing arrows of the two blocks of the first kind go into blocks of the second kind, there is one arrow of one of the two blocks of the second kind that has no place to go. So there has to be one arrow (and just one, in order not to have bidirectional arrows and spoil chirality) from a block of the first kind (we call it the first block) to the other block of the first kind (we call it the second block). From the second block will start exactly one arrow (since it is of the first kind), and this arrow will arrive in a block of the second kind; we call this block the third block. The remaining block will be the fourth one.\\
Up to now we have ordered the blocks and we have assigned the direction of three arrows:
\begin{center}
\ig[height=4.5 cm]{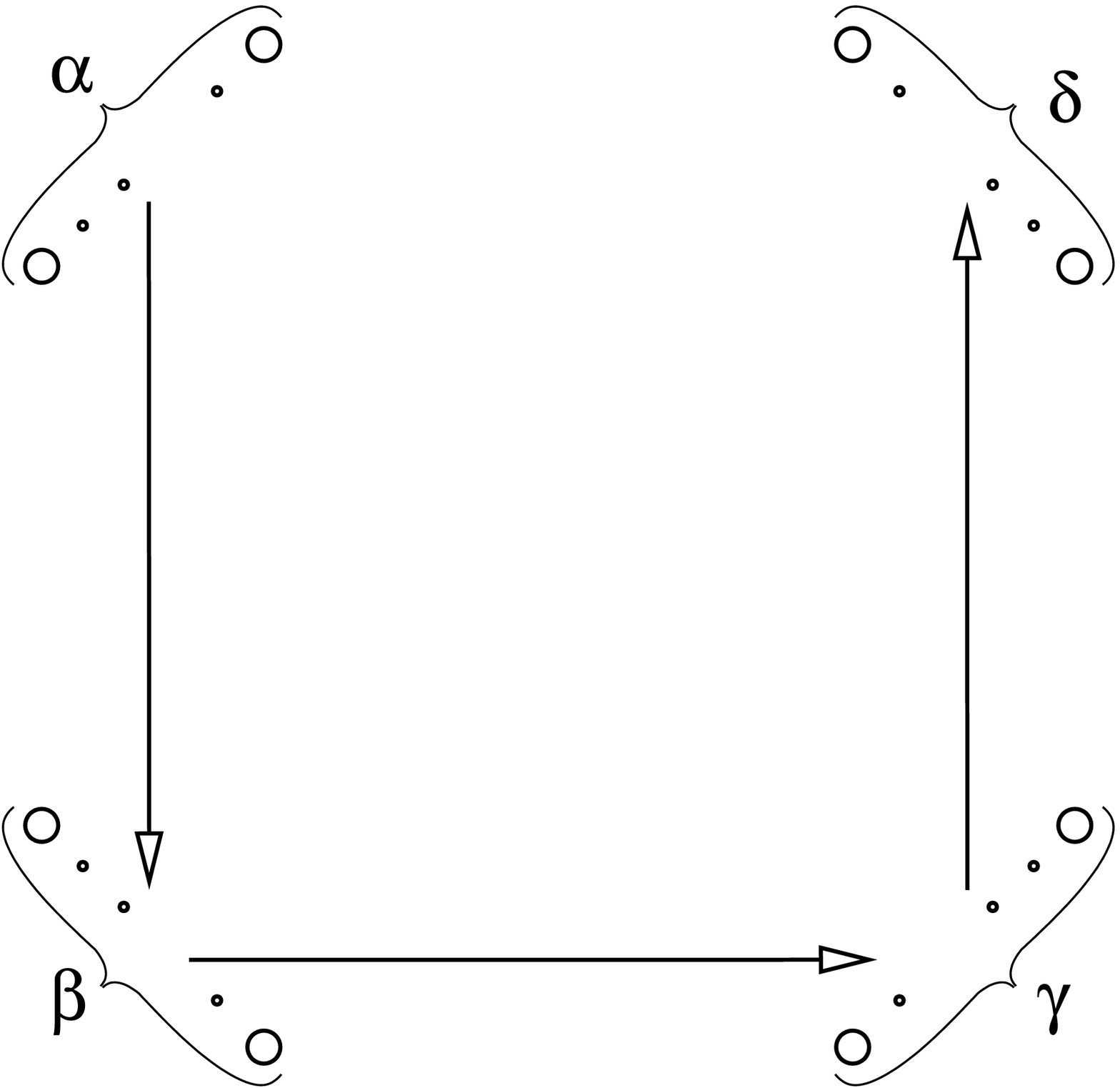}
\end{center}
The missing outgoing arrow from block-$3$ can go only in block-$1$, and the two arrows of block-$4$ must finally go in block-$1$ and block-$2$, completing the diagram as in figure \ref{figquiver_4block}.

In figure \ref{figquiver_4block} there are also the number of bifundamentals between nodes in different blocks (intersection numbers): $a$, $b$, ... , $f$. These are non-negative integer numbers.\\
We conclude that there is just one type of quiver diagram also in the case of four blocks.
In the case of 5 blocks (or nodes) this is not true anymore \cite{Herzog:2004}.

The quiver matrix $\mathcal{Q}_{i j}$ is thus an antisymmetric block-matrix, with blocks of dimension $\alpha \times \alpha$, $\alpha \times \beta$ etc. The first physical condition to be imposed is that for each node the ABJ gauge anomalies vanish, as in (\ref{ABJconstraint}): 
\beq\label{ABJconstraint4b} 
\sum_j \mathcal{Q}_{i j} N_j = 0
\;\;\;\; \text{for each}\;i, 
\eeq 
where $N_j$ is the rank of the $i^{th}$ node. The kernel of $\mathcal{Q}$ has dimension at least \mbox{$( dim[\mathcal{Q}] - 4 )$}, but we have to search the kernel for vectors satisfying the 4-block condition:
$$\big(  \underbrace{N_1, \ldots, N_1}_{\ag},\, \underbrace{N_2, \ldots, N_2}_{\bg},\,
     \underbrace{N_3, \ldots, N_3}_{\cg},\, \underbrace{N_4, \ldots, N_4}_{\dg}  \big)\,.$$
We have thus to look at a reduced antisymmetric quiver matrix:
\beq\label{reducedmatrix4b} q = \left(
\begin{array}{cccc}
    0   &  a   &  -b   & -c\\
   -a   &  0   &   d   & -e\\
    b   &  -d  &   0   &  f\\
    c   &  e   &  -f   &  0\\
\end{array}
\right) \;\,.\eeq

We recall that the minus signs mean that, for instance, $- b$
arrows go from block-$1$ to block-$3$, i.e. the arrows connecting
block-$1$ and $3$ go from $3$ to $1$, as in the Figure \ref{figquiver_4block}.

Canceling the gauge anomalies means finding a vector in the
kernel of $q$. In the 3-blocks chiral quivers there is always
exactly one such vector (we exclude the trivial case \mbox{$q =
0$}), since a $3 \times 3$ antisymmetric matrix has always rank
$2$. In the 4-block case this is not guaranteed: the vanishing of
the determinant has to be imposed: 
\beq\label{DETconstraint}
det(q) = (c \, d + b \, e - a \, f)^2 = 0\;. \eeq
 At this point we have a two-dimensional kernel, so, for any choice of $\ag, \bg,
\cg, \dg$ and of a singular matrix $q$, there are well defined
gauge theories (if the vector $(\ag N_1, \bg N_2, \cg N_3, \dg
N_4)$ lies in the kernel of $q$), however it is not guaranteed
that there is an interacting conformal fixed point.

Before going on to impose these last constraints we describe when
the $4$-block models reduce to $3$-block ones. The vanishing of
the gauge anomalies tells that, if, for instance, $a$ vanishes
also $b$ and $c$ have to vanish, and the model reduces to a
$3$-block model. The same happens for $d$ and $f$. Thus, if we
don't want to reduce to a $3$-block, we have to require also \beq a
> 0 \, , \;\; d > 0 \, , \;\; f > 0\;. \eeq
We remark that this condition is necessary in order to have a
$4$-block quiver, but it is not sufficient: it is possible that a
quiver satisfying the previous constraint can be Seiberg Dualised
to a $3$-block quiver.

\subsection{The superconformal conditions}
Now we arrive to the most restrictive contraints. We have to
impose that the beta functions of the gauge couplings and of the
superpotential couplings vanish.
We call the $r$-charges of the chiral superfields in the bifundamentals $r_a, r_b, ..., r_f$. The numerators of the exact beta functions for the $4$ gauge couplings are (as in (\ref{NSVZconstraint3a})):
\beq\label{NSVZconstraint} \left\{
\begin{array}{c}
N_1 + \frac{1}{2} \beta N_2 \, a (r_a - 1) + \frac{1}{2} \gamma N_3 \, b (r_b - 1) + \frac{1}{2} \delta N_4 \, c (r_c - 1) = 0\\

N_2 + \frac{1}{2} \alpha N_1 \, a (r_a - 1) + \frac{1}{2} \gamma N_3 \, d (r_d - 1) + \frac{1}{2} \delta N_4 \, e (r_e - 1) = 0\\

N_3 + \frac{1}{2} \alpha N_1 \, b (r_b - 1) + \frac{1}{2} \beta N_2 \, d (r_d - 1) + \frac{1}{2} \delta N_4 \, f (r_f - 1) = 0\\

N_4 + \frac{1}{2} \alpha N_1 \, c (r_c - 1) + \frac{1}{2} \beta N_2 \, e (r_e - 1) + \frac{1}{2} \gamma N_3 \, f (r_f - 1) = 0\,.
\end{array}
\right. \eeq 

There are also relations coming from the vanishing of the beta
functions of the superpotential couplings. From the quiver diagram
we see that there are three classes of gauge invariant operators
candidate to appear in the superpotential:
\begin{itemize}
\item a cubic term $X_{\alpha \beta}X_{\beta \gamma}X_{\gamma \alpha}$;
\item a cubic term $X_{\beta \gamma}X_{\gamma \delta}X_{\delta \beta}$;
\item a quartic term $X_{\alpha \beta}X_{\beta \gamma}X_{\gamma \delta}X_{\delta \alpha}$.
\end{itemize}
In order that each field enters in at least one term in the
superpotential it is necessary that all of these three terms appear
\footnote{This is precisely the requirement that will prevent us
from finding quivers with negative $r$-charges, of the type
discussed in \cite{Feng:2002kk}, \cite{Herzog:2004}.}. A chiral
operator is marginal at the interacting fixed point if its total
$r$-charge is $2$. So the $r$-charges have to satisfy the
following $3$ linear relations, analogous of (\ref{Wconstraint3b}):
\beq\label{Wconstraint} \left\{
\begin{array}{l}
r_a + r_d + r_b = 2\\
r_e + r_d + r_f = 2\\
r_a + r_d + r_f + r_c = 2\,.
\end{array}
\right. \eeq In total there are seven linear equations and six
variables. In order that a solution exists the integers $(a, b, c,
... , f)$ have to satisfy a constraint. It turns out that this
constraint is a diophantine equation, a higher order
generalization of the Markov Equation. Once this equation is
satisfied, we can reconstruct uniquely the $4$-vector of the ranks
in terms of the quiver matrix, and we can also find the $6$
$r$-charges. We note that of the $2$-dimensional kernel only one
vector $(N_1, N_2, N_3, N_4)$ gives rise to a superconformal
theory. The other vector takes the theory away from the conformal
point and can be used to study duality cascades of the
Klebanov-Strassler type \cite{klebanovstrassler}\cite{Franco:2003ea}\cite{Franco:2003ja}.

Instead of describing in detail the quite involved simultaneous solution of the
equations (\ref{ABJconstraint}), (\ref{DETconstraint}),
(\ref{NSVZconstraint}), (\ref{Wconstraint}) we just give the
results in the next paragraph and comment on the results.

\subsection{Results for the four blocks case}
The diophantine equation, written in term of only $( \alpha,
\beta, \gamma, \delta )$ and the quiver matrix, is a
generalization of the $3$-block one (\ref{dioph3blocks}):
\beq \label{diophantine4_blocks} \frac{a^2}{\cg \dg} +
\frac{b^2}{\bg \dg} + \frac{d^2}{\bg \dg} + \frac{d^2}{\ag \dg} +
\frac{e^2}{\ag \cg} + \frac{f^2}{\ag \bg} + \frac{a c e}{\cg} +
\frac{b c f}{\bg} = \frac{a b d}{\dg} + \frac{d e f}{\ag} + a c d
f \;\;, \eeq which has to be supplemented by the rank-$2$
constraint (\ref{DETconstraint}): 
\beq 
c d + b e = a f\;\;. 
\eeq
Equation (\ref{diophantine4_blocks}) is the generalization of the equation for 4 nodes: $\ag=\bg=\cg=\dg=1$ which was derived in \cite{Feng:2002kk}.
We note that the r.h.s. of (\ref{diophantine4_blocks}) is the sum of the monomials corresponding to the loops appearing in the superpotential, similarly to the $3$-block case.

The ranks of the gauge groups are given by the following formulae ($N_i = N \; x_i$):
\beq\label{ranks}
\left\{
\begin{array}{c}
\ag K^2 x_1^2 = \bg \cg \; d^2 +  \bg \dg \; e^2 + \cg \dg \; f^2 - \bg \cg \dg \; d e f\\
\bg K^2 x_2^2 = \ag \cg \; b^2 +  \ag \dg \; c^2 + \cg \dg \; f^2 + \ag \cg \dg \; b c f\\
\cg K^2 x_3^2 = \ag \bg \; a^2 +  \ag \dg \; c^2 + \bg \dg \; e^2 + \ag \bg \dg \; a c e\\
\dg K^2 x_4^2 = \ag \bg \; a^2 +  \ag \cg \; b^2 + \bg \cg \; d^2 - \ag \bg \cg \; a b d\,.\\
\end{array}
\right.
\eeq
$K^2$ is a priori a rational number such that the four ranks $x_i$ have no common factors. Also in this case it turns out that it is always an integer depending only on the number of nodes. This fact can be seen as a consequence of algebro-geometric results for del Pezzo quivers and of the fact that all the new quivers are obtained by del Pezzo quivers shrinking one or two blocks of $4$ nodes. The relation is as in (\ref{KSQ-nodes}): $K^2$ is given by $12$ minus the number of nodes for a del Pezzo quiver and by $9$ minus the number of nodes for a del Pezzo quiver where $4$ nodes have been shrunk.

It is also possible to find a general formula for the $r$-charges:
\beq\label{rcharges}
\left\{
\begin{array}{c}
 K^2 r_a = \frac{2 a}{x_1 x_2}\\
 K^2 r_d = \frac{2 d}{x_2 x_3}\\
 K^2 r_f = \frac{2 f}{x_3 x_4}\,.\\
\end{array}
\right. 
\eeq 
$r_b, r_c, r_e$ can be obtained combining this relations with the superpotential constraints (\ref{Wconstraint}).

We note that the results (\ref{diophantine4_blocks}), (\ref{ranks}) and (\ref{rcharges}) are in agreement with general formulae derived in \cite{Herzog:2003zc} studying properties of exceptional collections of sheaves over del Pezzo surfaces. The implication is that the properties of del Pezzo quivers are expected to be valid for any chiral quiver.

\subsubsection*{A check: reduction to three blocks}
The formulae of the last subsection can be specified to the case of three blocks. In order to do this, it is enough to put $x_4 = 0$ or $x_1 = 0$. It is not possible to ask for the vanishing of $x_2$ or $x_3$, since in these cases equations (\ref{ranks}) and (\ref{diophantine4_blocks}) would force the entire quiver matrix to vanish. This is consistent with the fact that if the block $\bg$ or $\cg$ disappears the quiver diagram reduces to a non-oriented triangle.

For definitess we take $x_4 = 0$ and $c = e = f = 0$. The equations for the ranks become:
\beq\label{ranks3b}
\left\{
\begin{array}{l}
\ag K^2 x_1^2 = \bg \cg \; d^2 \\
\bg K^2 x_2^2 = \ag \cg \; b^2 \\
\cg K^2 x_3^2 = \ag \bg \; a^2 \\
0 = \ag \bg \; a^2 +  \ag \cg \; b^2 + \bg \cg \; d^2 - \ag \bg \cg \; a b d\,.\\
\end{array}
\right. \eeq 
The rank-$2$ constraint (\ref{DETconstraint}) is identically satisfied (a non vanishing antisymmetric $3 \, x \, 3$ matrix has always a one dimensional kernel) and the diophantine equation coincides with the fourth equation of (\ref{ranks3b}). In summary (\ref{ranks3b}) coincides exactly with the results obtained in the 3-block case and we have a check of the validity of the formulae given in the previous subsection.

\section{Conclusions}

In this paper general results concerning superconformal quivers are found.

We showed that for all these theories the relation $c = a$ is satisfied. This suggest that every superconformal quiver can be geometrically engineered as the low energy theory arising on D3 branes probing a singular Calabi-Yau threefold. 

Furthermore we defined an operation on superconformal quivers that changes the quiver but leaves the superconformal constraints satisfied. This ``shirinking'' procedure decreases the number of nodes, and can be applied any time the original quiver has a block of, at least, four nodes.

Restricting the attention to the case of completely chiral quivers, we showed that all models that admit a $3$ or $4$ block structure are the well known del Pezzo quivers or arise from shrinking one or two times a del Pezzo quiver.

\vspace{0.2 cm}

There are thus two main problems that are left open.

First it would be nice to complete the classification of rank $2$ chiral quivers. Our results for $3$ and $4$ blocks suggest that all these models are del Pezzo quivers or arise from shrinking a del Pezzo quiver. This would imply that the maximum number of nodes is precisely $11$ and that all the physical requirements discussed in section \ref{threeblock} (unitarity and asymptotic freedom) are always satisfied.

The second problem concerns understanding if there is a string theoretical description of the procedure of shrinking. One possibility is that these new theories arise on D3 branes on vanishing del Pezzo surfaces as the del Pezzo quivers, but with Discrete Torsion turned on. Such backgrounds are notoriously difficult to analize, however it should be possible to gain some insights on this problem directly from the gauge theory. For instance imposing SU(2) symmetry in the shrunk models the superpotential is highly constrained: it could be possible to understand which are the exactly marginal superpotentials and to reconstruct the moduli space of vacua.

Independently from the knowledge of the superpotential, some information could be gained from the study of baryons and baryonic symmetries, along the lines of \cite{Intriligator:2003wr, Herzog:2003wt, Herzog:2003dj}. It would also be nice to understand if there are hidden global symmetries as the ones found for del Pezzo quivers in \cite{Franco:2004}.


\section*{Acknowledgements}
We would like to thank Damiano Anselmi, Guido Festuccia, Sebastian Franco and Pavlos Kazakopoulos for useful and enjoyable conversations. S.~B. also wishes to aknowledge the kind hospitality of CTP, where a large part of this work has been done.

This research is supported in part by the CTP and the LNS of MIT.
Further support is granted from the U.S. Department of Energy
under cooperative agreements $\#$DE-FC02-94ER40818. A.~H. is also
supported by the Reed Fund Award and a DOE OJI award. S.~B. is
also supported by INFN-MIT ``Bruno Rossi" Exchange Program.

\appendix

\vspace{1 cm}

\section{Classification of three-block diophantine equations}\label{classification}

In section \ref{threeblock} it has been shown that the three-block models are classified by the following equation:
\beq
\frac{a^2}{\alpha}+\frac{b^2}{\beta}+\frac{c^2}{\gamma} = a b c
\label{markov_3block}
\eeq

We want to find all the solutions to this equations, modulo
Seiberg Dualities. Given a particular solution we can Seiberg
dualize it and obtain another solutions in the Duality Tree. (In
the case $\alpha =\beta =\gamma = 1$ this is the so called Markov
Tree). As remarked in section \ref{dualitytrees} it is convenient
to restrict the search to the roots of the trees. We will see that
there are only a finite number of ``root'' solutions to the
generalized Markov equation. Equivalently there are only a finite
number of triples $(\alpha, \beta, \gamma)$ that admit a solution
for (\ref{markov_3block}).

In order to simplify the search, it is convenient to decrease the
number of the variables (now we have the six integer variables $a,
b, c, \alpha, \beta, \gamma$, from which we can easily obtain also
the other three integer $x, y, z$ and the rational number $K^2$)
and work with just three new integer variables: 
\beq \label{defin}
\left\{
\begin{array}{l}
X := a^2 \beta \gamma    \\
Y := b^2 \alpha \gamma   \\
Z := c^2 \alpha \beta
\end{array}
\right.
\eeq
In term of this variables (\ref{markov_3block}) becomes
\beq\label{3block}
X + Y + Z = \sqrt{X Y Z} \;\,.
\eeq

By construction each solution of (\ref{markov_3block}) gives rise to a solution of (\ref{3block}); thus, once we find all the integer solution of (\ref{3block}) we can find all the solutions of (\ref{markov_3block}), and this last step is very easy, since the defining equations for $X, Y, Z$ are simple. Equation (\ref{3block}) is satisfied by an infinite number of triples of integer, however also these triples can be related by ``Seiberg Duality'', so they can be organized in ``duality trees'': we will see that there are just \emph{four} duality trees for (\ref{3block}).

\subsection*{$\;\;\;$ Study of the equation $\;\;\;\; X + Y + Z = \sqrt{X Y Z}$}
Also for this equation, given a particular integer solution, we
can find an infinite set of solutions ``dualising'' one of the
integers of the triple. The reason is that equation (\ref{3block})
is quadratic in the variable $\sqrt{Z}$ (or $\sqrt{Y}$ or
$\sqrt{X}$).
The dual can be found directly writing (\ref{3block}) in the form
\beq\label{interm}
X + Y = \sqrt{Z} (\sqrt{X Y} - \sqrt{Z}) := \sqrt{Z}\sqrt{\tilde{Z}}
\eeq
Where we have defined the dual of $Z$ to be $\tilde{Z}$:
\beq
\tilde{Z} = (\sqrt{X Y} - \sqrt{Z})^2 = X Y + Z - 2 \sqrt{X Y Z}
\eeq
Since $\sqrt{X Y Z}$ is an integer also $\tilde{Z}$ will be an integer.

In order to classify all duality trees it is enough to classify their minimal
solutions. Also in this case we define a solution to be minimal
(the ``root'' of the tree) if $X$, $Y$ and $Z$ increase under dualization.
Under the assumption $X \leq Y \leq Z$ it is enough to ask that $Z$ increases
under the duality, that is $\tilde{Z}$ is greater than $Z$.\\
(\ref{interm}) implies that a minimal solution ($\tilde{Z} \geq Z$) satisfies
\beq\label{s.d.}
\tilde{Z} \geq Z \;\;\rightarrow\;\; X + Y = \sqrt{Z} \sqrt{\tilde{Z}} \geq \sqrt{Z} \sqrt{Z} = Z\;.
\eeq
Modulo trivial reorderings of the triple $( X, Y, Z )$, the problem is thus reduced to the search of all the integer solutions of the following system:
\beq \label{sistem}
\Bigg\{
\begin{array}{l}
X + Y + Z = \sqrt{X Y Z}\\
X \leq Y \leq Z \leq X + Y\;\,.
\end{array}
\eeq
The solutions of this system are exactly the minimal solution of
equation (\ref{3block}). We claim that the only positive-integer solutions
of (\ref{sistem}) are the four listed in table \ref{4sols}.

\begin{table}[!h]
\begin{center}
$$\begin{array}{|c|c|c|}  \hline
  \hs  X \hs & \hs  Y \hs  & \hs  Z \hs \\ \hline\hline
    5      &  20  &  25  \\\hline
    6      &  12  &  18  \\\hline
    8      &  8   &  16  \\\hline
    9      &  9   &  9    \\  \hline
\end{array}$$
\caption{The four minimal solutions of the diophantine equation $X + Y + Z = \sqrt{X Y Z}$.}
\label{4sols}
\end{center}
\end{table}

In order to prove this fact it is possible to proceed as follow.

First of all it is easy to show that there is a lower bound for
$X$ stricter than $0$. Solving the equation (\ref{3block}) for
$\sqrt{Z}$ gives
\beq\label{solution}
\sqrt{Z} = \frac{1}{2}\left(\sqrt{X Y} \pm \sqrt{X Y - 4 X - 4 Y}\right)
\eeq
In order to have solutions it is necessary that the expression
inside the square root is non-negative:
\beq
X Y - 4 X - 4 Y \geq 0 \;\;\Rightarrow\;\; Y \, (X - 4) \geq 4 \, X > 0 
\;\;\Rightarrow\;\; X > 4\;.
\eeq
Since we search integers the last restriction is $X \geq 5$.

Second we look for a higher bound for $X$. In order to find it we
can study the system (\ref{sistem}) for real values of $X, Y, Z$.
In particular we fix a value of $X \geq 5$ and plot in the plane
$( Y, Z )$ the solution (\ref{solution}) and the linear constraints of (\ref{sistem}):

\begin{figure}[!h]
\begin{center}
\ig[height=7 cm]{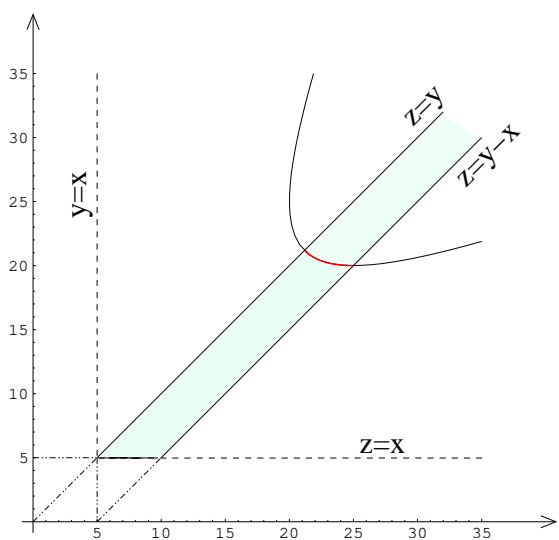}
\end{center}\caption{Intersection between the solution to equation $X + Y + Z = \sqrt{X Y Z}$ and the region $X \leq Y \leq Z \leq X + Y$ for a fixed value of $X$. In this figure $X = 5$.}\label{intersection}
\end{figure}

\begin{figure}[!h]
\begin{center}
\ig[height=4.5 cm]{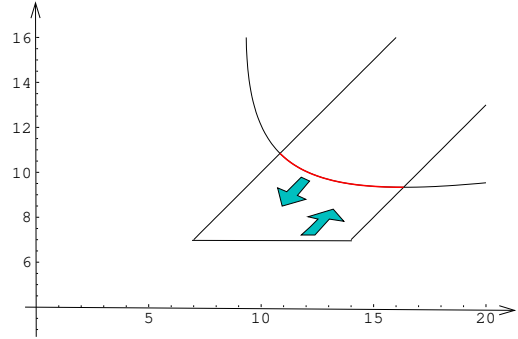}
\end{center}\caption{Effect of the increase of the value of $X$. There is a maximum value of $X$ over which the solution to $X + Y + Z = \sqrt{X Y Z}$ doesn't pass over the region containing the minimal solutions.}\label{moving}
\end{figure}

From the graph it is clear that there could be solutions to the system (\ref{sistem}) only on a finite
piece of the curved line representing (\ref{solution}).\\
Let's consider what happens increasing the value of $X$: the curved line corresponding to the solution (\ref{solution}) moves in the left-down direction, while the allowed region (delimited by the linear constraints) move in the right-up direction (Figure \ref{moving}). 

The conclusion is that there is a maximum value of $X$ allowing (real)
solutions of (\ref{sistem}). This maximum value is reached when the curved line meets the point of intersection of the lines $Z = X$ and $Z = Y$. In this case we have $X = Y = Z$ and the equation (\ref{3block}) gives:
$$3 X = \sqrt{X^3} \;\; \Rightarrow \;\; X = Y = Z = 9\;\,.$$
Thus at this extremal point there is an integer solution. \\
Assuming now that $X$ is integer, the search for duality trees is restricted to four cases: $5 \leq X \leq 8$.

Now the classification is easy but still quite long; in order to abbreviate it we show that, in all of the four cases, the fact that the square root $\sqrt{X Y Z}$ has to be an integer implies that \emph{both $Y$ and $Z$ must be multiples of $X$} \footnote{We point out that this property does not hold in general for non minimal solution, if the restriction $X \leq Y \leq Z \leq X + Y$ is relaxed. What is true is, for instance, that all the triples in the duality tree with root $( 5, 20, 25 )$ are made of integer multiples of $5$. This can be seen easily from (\ref{s.d.}) using the property just stated, that we are going to show.}.
\begin{quotation}
\begin{small}
\noindent If $X = 5$ (\ref{3block}) becomes
$$5 + Y + Z = \sqrt{5 Y Z}$$
so one of the two integers $Y$ and $Z$ has to be a multiple of $5$, else $5 Y Z$ contains the factor $5$ just one time, and cannot be a perfect square. Then we have
$$5 + 5 k + h = 5 \sqrt{k h}\;,$$
so $5$ has to divide $h$: both $Y$ and $Z$ have to be multiples of $5$.\\
Exactly the same argument works for $X = 7$ and also if all the prime factors in $X$ appear just one time ($X = 6$).\\
If $X = 8$ (\ref{3block}) reads
$$8 + Y + Z = 2 \sqrt{2 Y Z}\;.$$
One of $Y$ or $Z$ has to be a multiple of $2$; this implies like above that also the other one does. We thus divide the equation by $2$:
$$4 + Y' + Z' =  2 \sqrt{2 Y' Z'}$$
Now either $Y'$ or $Z'$ has to be even, then also the other does. Dividing again by $2$:
$$2 + Y'' + Z'' =  \sqrt{2 Y'' Z''}$$
The same argument again shows that both $X$ and $Y$ must be integer multiple of $8$.\\
\end{small}
\end{quotation}
\noindent The consequence is that for $5 \leq X \leq 8$ the system (\ref{sistem}) can be written as ($r = \frac{Y}{X}$, $s =
\frac{Z}{X}$) \beq\label{sistem2} \Bigg\{
\begin{array}{l}
1 + r + s = \sqrt{X r s}\\
1 \leq r \leq s \leq r + 1
\end{array}
\eeq with integer $r$ and $s$. The case $r = s$ is excluded because $X$ isn't a perfect square in the range $5 ... 8$. So it must be $s = r + 1$ and 
\beq \label{final} 
    2 (r + 1) = \sqrt{X r (r+1)} \;\; \Rightarrow \;\; 4 (r + 1) = X r 
    \;\; \Rightarrow \;\; r = \frac{Y}{X} = \frac{4}{X - 4} 
\eeq 
The last formula gives finally the first three solutions of table (\ref{4sols}), since $Y$ and $Z$ are integer except for $X = 7$.

\subsection*{Reconstruction of the solutions of the three-block Markov equations}
Now we have to reconstruct the three block quivers, starting from the four triples of Table \ref{4sols}. This is a bit long but  straightforward, so we discuss just the more involved example: $(X, Y, Z) = (8, 8, 16)$. This is precisely the case where we find quivers not directly related to del Pezzo surfaces. We will see that the difference can be traced to the value of $K^2$.

From the definitions given in (\ref{defin}) we first have to find  $\{ a, b, c \,;\, \alpha, \beta, \gamma \}$ such that:
\beq \label{defin8816}
\left\{
\begin{array}{l}
\;8 = a^2 \, \beta \gamma    \\
\;8 = b^2 \, \alpha \gamma   \\
 16 = c^2 \, \alpha \beta \; .
\end{array}
\right. \eeq
Since there is a symmetry between the first two equations, we can ask for instance $\ag \geq \bg$. The third equation says that $c$ can be $1$, $2$ or $4$.


 \begin{footnotesize}

 $$
 \begin{array}{l c c c r}
   c = 1
    &
   \longrightarrow
    &
   \left\{\begin{array}{l}
   \;8 = a^2 \beta \gamma    \\
   \;8 = b^2 \alpha \gamma   \\
    16 = \alpha \beta
   \end{array}\right.
    &
   \setlength{\unitlength}{1 mm}
   \begin{picture}(18, 0)\put(3,0){\vector(2,1){15}}\put(3,0){\vector(2,-1){15}}\end{picture}
    &
  \begin{array}{c c c}
   \hspace{-0.5cm}
    \left\{ \begin{array}{l}
      \ag = \bg = 4  \\
      2 = a^2  \cg  \\
      2 = b^2  \cg
   \end{array}\right.
                         &       \longrightarrow         &
                                                           \left\{ \begin{array}{l}
                                                           \ag = \bg = 4  \;\;\;\; \cg = 2\\
                                                            a = b = 1
                                                            \end{array}\right.    \\
                        &                                &                        \\
   \;\left\{ \begin{array}{l}
     \ag = 8 \;\;\;\; \bg = 2  \\
     4 = a^2  \cg  \\
     1 = b^2  \cg
   \end{array} \right.
                         &       \longrightarrow         &
                                                           \left\{ \begin{array}{l}
                                                           \ag = 8 \;\;\;\; \bg = 2  \;\;\;\; \cg = 1\\
                                                            a  = 2 \;\;\;\; b = 1
                                                            \end{array}\right.   \\
  \end{array}
 \end{array}
 $$

\vspace{0.3 cm}

 $$
 \begin{array}{l c c c r}
   c = 2
    &
   \longrightarrow
    &
   \left\{\begin{array}{l}
     8 = a^2 \beta \gamma    \\
     8 = b^2 \alpha \gamma   \\
     4 = \alpha \beta
   \end{array}\right.
    &
   \setlength{\unitlength}{1 mm}
   \begin{picture}(18, 0)\put(3,0){\vector(4,3){15}}\put(3,0){\vector(2,-1){15}}\end{picture}
    &
    \begin{array}{c c c}
     \left\{ \begin{array}{l}
       \ag = 4 \;\;\;\;  \bg = 1  \\
       8 = a^2  \cg  \\
       2 = b^2  \cg
     \end{array}\right.
                         &       \longrightarrow         &
                                                           \left\{ \begin{array}{l}
                                                           \ag = 4 \;\;\;\; \bg = 1  \;\;\;\; \cg = 2\\
                                                            a = 2  \;\;\;\;  b = 1
                                                            \end{array}\right.    \\
                        &                                &                          \\
     \hspace{-0.6 cm}
      \left\{ \begin{array}{l}
       \ag = \bg = 2  \\
       4 = a^2  \gamma  \\
       4 = b^2  \gamma
     \end{array} \right.
              &       \setlength{\unitlength}{1 mm}
   \begin{picture}(15, 0)\put(0,0){\vector(2,1){15}}\put(0,0){\vector(2,-1){15}}\end{picture}        &
                                                           \begin{array}{l}
                                                               \left\{ \begin{array}{l}
                                                                 \ag = \bg = 2  \;\;\;\; \cg = 1\\
                                                                 a  =  b = 2
                                                               \end{array}\right.   \\ 
                                                                        \\ 
                                                                        
                                                               \left\{ \begin{array}{l}
                                                                 \ag = \bg = 2  \;\;\;\; \cg = 4\\
                                                                   a  = 1 \;\;\;\; b = 1
                                                               \end{array}\right.   \\
                                  \end{array}
                                                         \\
  \end{array}
 \end{array}
 $$

\vspace{0.3 cm}
 $$
 \begin{array}{l c c c c c r}
  c = 4
    &
 \;\; \longrightarrow \;\;
    &
   \left\{\begin{array}{l}
            8 = a^2 \beta \gamma    \\
            8 = b^2 \alpha \gamma   \\
              1 = \alpha \beta
          \end{array}
   \right.
    &
  \;\; \longrightarrow \;\;
    &
   \begin{array}{c c c}
    \left\{ \begin{array}{l}
             \ag = \bg = 1  \\
               8 = a^2  \gamma  \\
               8 = b^2  \gamma
            \end{array}
    \right.
     &
   \setlength{\unitlength}{1 mm}
   \begin{picture}(15, 0)\put(0,0){\vector(2,1){15}}\put(0,0){\vector(2,-1){15}}\end{picture}
     &
     \begin{array}{l}
          \left\{ \begin{array}{l}
                    \ag = \bg = 1 \;\;\;\; \cg = 2\\
                     a =  b = 2
                  \end{array}\right.   \\  \\
          \left\{ \begin{array}{l}
                  \ag = \bg = 1  \;\;\;\; \cg = 8\\
                   a  =  b = 1
                  \end{array}\right.   \\
     \end{array}
          \\
   \end{array}
 \end{array}
 $$

\end{footnotesize}

The other three cases are shorter, we just give the results in the table.

The values of $K^2$ and of $x, y, z$ are determined by formulae (\ref{Ksq})
$$\left\{
\begin{array}{l}
K^2 x^2 = a^2 \frac{\bg \cg}{\ag}\\
K^2 y^2 = b^2 \frac{\ag \cg}{\bg}\\
K^2 z^2 = c^2 \frac{\ag \bg}{\cg}
\end{array}
\right.$$
 imposing that the greatest common divisor of $x, y, z$ is $1$. Formulae (\ref{r-charges})
$$\left\{
\begin{array}{c}
r_a  = \frac{2 a}{\ag b c}\\
r_b  = \frac{2 b}{\bg a c}\\
r_c  = \frac{2 c}{\cg a b}
\end{array}\right.$$
give the three $r$-charges. We note that there is a one-to-one correspondence
between the $r$-charges of the minimal solution of a given duality
tree and the four families of Markov type equations. In the last
column there is the del Pezzo surface related to the quiver. We dubbed
the two shrunk models ``sh-del Pezzo''.\\

\begin{small}
\begin{tabular}{|c|ccc|ccc|ccc|c|ccc|c|}
\hline
$(X ,Y ,Z)$  &$\ag$&$\bg$&$\cg$  &  a & b & c &    x & y & z   &$K^2$ &  $r_a$ & $r_b$ & $r_c$ & surface\\
\hline
\hline
            &  1  &  1  &   1    &  3 & 3 & 3 &    1 & 1 & 1    & 9  &   2/3 & 2/3 & 2/3   &  $dP_0$ \\
\cline{2-15}
 $(9 ,9 ,9)$   &  3  &  3  &   3 &  1 & 1 & 1 &    1 & 1 & 1    & 3  &   2/3 & 2/3 & 2/3   &  $dP_6$ \\
\cline{2-15}
               &  1  &  1  &   9 &  1 & 1 & 3 &    3 & 3 & 1    & 1  &   2/3 & 2/3 & 2/3   &  $dP_8$ \\
\hline
\hline
               &  1  &  1  &   2 &  2 & 2 & 4 &    1 & 1 & 1    & 8  &   1/2 & 1/2 &  1    &  $F_0$ \\
\cline{2-15}
               &  2  &  2  &   4 &  1 & 1 & 2 &    1 & 1 & 1    & 4  &   1/2 & 1/2 &  1    &  $dP_5$ \\
\cline{2-15}
               &  4  &  4  &   2 &  1 & 1 & 1 &    1 & 1 & 2    & 2  &   1/2 & 1/2 &  1    &  $dP_7$ \\
\cline{2-15}
  $(8, 8, 16)$ &  1  &  1  &   8 &  1 & 1 & 4 &    2 & 2 & 1    & 2  &   1/2 & 1/2 &  1    &  $dP_7$ \\
\cline{2-15}
               &  8  &  2  &   1 &  2 & 1 & 1 &    1 & 2 & 4    & 1  &   1/2 & 1/2 &  1    &  $dP_8$ \\
\cline{2-15}
               &  2  &  2  &   1 &  2 & 2 & 2 &    1 & 1 & 2    & 4  &   1/2 & 1/2 &  1    & $sh\,dP_5$ \\
\cline{2-15}
               &  2  &  1  &   4 &  2 & 1 & 2 &    2 & 2 & 1    & 2  &   1/2 & 1/2 &  1    & $sh\,dP_7$ \\
\hline
\hline
               &  1  &  2  &   3 &  1 & 2 & 3 &    1 & 1 & 1    & 6  &  1/3 & 2/3 &  1     &  $dP_3$ \\
\cline{2-15}
 \rb{0 cm}{$(6, 12, 18)$}
               &  2  &  1  &   6 &  1 & 1 & 3 &    1 & 2 & 1    & 3  &  1/3 & 2/3 &  1     &  $dP_6$ \\
\cline{2-15}
               &  3  &  1  &   6 &  1 & 1 & 2 &    1 & 3 & 1    & 2  &  1/3 & 2/3 &  1     &  $dP_7$ \\
\cline{2-15}
               &  2  &  3  &   6 &  1 & 1 & 1 &    3 & 2 & 1    & 1  &   1/3 & 2/3 &  1    &  $dP_8$ \\
\hline
\hline
 $(5, 20, 25)$ &  1  &  1  &   5 &  1 & 2 & 5 &    1 & 2 & 1    & 5  &  1/5 & 4/5 &  1    &  $dP_4$ \\
\cline{2-15}
               &  5  &  5  &   1 &  1 & 2 & 1 &    1 & 2 & 5    & 1  &  1/5 & 4/5 &  1    &  $dP_8$ \\
\hline
\end{tabular}
\end{small}

\vspace{1 cm}

From this table it is possible to verify the relations (\ref{KSQ-nodes})
\beq
\left\{
\begin{array}{lc}
K^2  = 12 - (\ag + \bg +\cg)  &  \text{for del Pezzo quivers}\\
K^2  = 9  - (\ag + \bg +\cg)  &  \text{for the ``new'' quivers}
\end{array}
\right.
\eeq
relating $K^2$ to the total number of nodes.

\newpage

 \bibliographystyle{JHEP}

\end{document}